\documentclass[article,structabstract]{aa}

\usepackage{xspace}
\usepackage{graphicx}
\usepackage{amsmath}
\usepackage{txfonts}
\usepackage{natbib}
\bibpunct{(}{)}{;}{a}{}{,} 
\newcommand{\um}{\ensuremath{\rm \mu m}\xspace}
\newcommand{\e}[1]{\ensuremath{\rm ~\times~10^{#1}}\xspace}
\newcommand{\ten}[1]{\ensuremath{\rm 10^{#1}}\xspace}

\newcommand{\alfaturb}{\ensuremath{\alpha_{\rm turb}}\xspace}
\newcommand{\dtg}{dust-to-gas\xspace}

\begin{document}

\title{Probing the turbulent mixing strength in protoplanetary disks across the stellar mass range: no significant variations.}

\authorrunning{Mulders \& Dominik}
\titlerunning{Probing turbulent mixing across the stellar mass range}

\author{Gijs~D.~Mulders\inst{1,2} \and Carsten~Dominik\inst{1,3}}

\institute{
Astronomical Institute ``Anton Pannekoek'', University of Amsterdam, PO Box 94249, 1090 GE Amsterdam, The Netherlands
\and
SRON Netherlands Institute for Space Research, PO Box 800, 9700 AV,
Groningen, The Netherlands
\and
Department of Astrophysics/IMAPP, Radboud University Nijmegen,
P.O. Box 9010 6500 GL Nijmegen The Netherlands
}

\offprints{Gijs Mulders, \email{\bf{mulders@uva.nl}}}

\abstract{
Dust settling and grain growth are the first steps in the planet-formation process in protoplanetary disks. These disks are observed around stars with different spectral types, and there are indications that the disks around lower mass stars are significantly flatter, which could indicate that they settle and evolve faster, or in a different way. 
}{
  We aim to test this assumption by modeling the median spectral energy distributions (SEDs) of three samples of protoplanetary disks: around Herbig stars, T Tauri stars and brown dwarfs. We focus on the turbulent mixing strength to avoid a strong observational bias from disk and stellar properties that depend on stellar mass.
}{
  We generated SEDs with the radiative transfer code MCMax, using a hydrostatic disk structure and settling the dust in a self-consistent way with the $\alpha$-prescription to probe the turbulent mixing strength.
}{
  We are able to fit all three samples with a disk with the same input parameters, scaling the inner edge to the dust evaporation radius and disk mass to millimeter photometry. The Herbig stars require a special treatment for the inner rim regions, while the T-Tauri stars require viscous heating, and the brown dwarfs lack a good estimate of the disk mass because only few millimeter detections exist.
}{
  We find that the turbulent mixing strength does not vary across the stellar mass range for a fixed grain size distribution and gas-to-dust ratio. Regions with the same temperature have a self-similar vertical structure independent of stellar mass, but regions at the same distance from the central star appear more settled in disks around lower mass stars. We find a relatively low turbulent mixing strength of $\alpha = 10^{-4}$ for a standard grain size distribution, but our results are also consistent with $\alpha = 10^{-2}$ for a grain size distribution with fewer small grains or a lower gas-to-dust ratio.
}

\maketitle

\section{Introduction}
Protoplanetary disks are the main sites of planet formation. Within them, the small sub-micron sized dust grains grow to millimeter and centimeter sizes and settle to the midplane, where they eventually form kilometer-sized planetesimals and proto-planets that are the building blocks of planetary systems. How the dust grows and settles depends not only on the grain size and gas density, but also on the turbulent mixing strength of the gas that mixes small grains back up into the disk atmosphere and makes them collide in the midplane. Without turbulent mixing, all disks would be flat and dust would grow slower. Additionally, turbulent mixing is an important driver for the viscous evolution of the disk, for radial mixing of dust, and it may also play an important role in the later stages of planet formation.

Although turbulent mixing is fundamental for our understanding of disk evolution and planet formation, the cause of disk turbulence is not completely understood \citep[e.g.][]{1998AIPC..431...79B,2011ARA&A..49..195A}. Turbulent mixing in disks is often implemented using the $\alpha$ prescription from \cite{1973A&A....24..337S} and \cite{1981ARA&A..19..137P}, but the value of the turbulent mixing strength \alfaturb is hard to predict from theory. Although it could be empirically derived from line-broadening with future observatories such as ALMA, this is challenging with current facilities \citep{2011ApJ...727...85H}. The value typically used and inferred from disk models trying to match accretion rates is $\alpha_{\rm viscous}$=0.01 \citep[e.g.][]{1998ApJ...495..385H}. Comparison of steady-state accretion disk models to millimeter images yields $\alpha_{\rm viscous}= 0.5...10^{-4}$ \citep{2009ApJ...701..260I}.

The turbulent mixing strength manifests itself in the degree of dust settling. When dust grains grow to larger sizes, they decouple from the gas and start settling toward the midplane. At what size and height they decouple depends not only on the gas density and temperature but also on the turbulent mixing strength: a lower turbulent mixing strength leads to flatter disks \citep[e.g.][]{2004A&A...421.1075D}. As dust decouples, its relative velocity also increases, which leads to fragmentation, providing a natural stop to grain growth and replenishing the small dust grain population that is then mixed up into the disk surface \citep{2005A&A...434..971D}. 

There is by now a wealth of observational evidence supporting grain growth and dust settling in protoplanetary disks. Most notably, mid-infrared spectroscopic observations have shown that grains in the warm surface layers have grown beyond the size of a micron \citep[e.g.][]{2009arXiv0911.1010H}. In addition, the spectral index at (sub)millimeter and centimeter wavelengths shows that grains grow up to centimeter sizes in the cold disk midplane \citep[e.g.][]{1991ApJ...381..250B,1994MNRAS.267..361M}. This picture has been confirmed with hydrostatic radiative transfer models by \cite{2006ApJ...638..314D} and \cite{2005ApJ...628L..65F} for T Tauri stars, who require a depletion of small dust grains at the disk surface and a population of big grains in the disk midplane. Without grain growth and dust settling, hydrostatic disk models overpredict observed far-infrared fluxes.

Interestingly enough, this seems not to be the case for Herbig Ae and Be stars, the intermediate mass counterparts of T Tauri stars. Hydrostatic disk models do well in explaining far-infrared fluxes without the need to deplete the upper layers of the disk \citep{2003A&A...398..607D, 2004A&A...417..159D,2009A&A...502L..17A}, especially in Meeus group I sources\footnote{We note that for group II sources, some settling may be required to settle the outer disk into the inner disk shadow.}. However, the (sub) millimeter slopes show that a population of millimeter-sized grains exist in these disks as well \citep[e.g.][]{2004A&A...422..621A}, indicating that the grain-size population is not that different from T Tauri stars. This opens up the possibility that these disks are less settled, and that disk properties - in particular the turbulent mixing strength - could vary strongly as a function of stellar mass. The recent discovery that disks around brown dwarfs appear to be more settled than T Tauri stars fits well into this picture \citep[e.g.][]{2006AJ....131.1574L,2010ApJ...720.1668S}, although this is not always found \citep{2011ApJS..195....3F}, and indicates that the observed degree of settling could be inversely proportional to stellar mass.

However, the degree of settling does not directly reflect the turbulent mixing strength. Disks around T Tauri stars, Herbig stars and brown dwarfs vary in disk mass \citep[e.g. ][]{1990AJ.....99..924B, 2006ApJ...645.1498S, 2009ApJ...692.1609V} and stellar mass, and observations at the same wavelength probe different radial regions in the disk because of the changing stellar luminosity. Comparing the degree of settling using a fixed height or a lower scale height for bigger grains \citep[e.g. ][]{2006ApJ...638..314D, 2011A&A...527A..27S} is therefore less suitable for comparison along the stellar mass range. Radiative transfer models that describe dust settling using turbulent mixing exist \citep{2004A&A...421.1075D,2010MNRAS.401..143H}, but have not been used to compare to observations directly.

To constrain the turbulent mixing strength across the stellar mass range, we focused primarily on median spectral energy distributions (SEDs). We did this because measuring the degree of settling in individual disks is not straightforward because of a number of degeneracies in disk modeling \citep{2011A&A...527A..27S}, and requires detailed multiwavelength modeling, including scattered-light images \citep{2008A&A...489..633P}, but conclusions can be drawn from larger samples \citep{2005ApJ...628L..65F,2006ApJS..165..568F}. They are available for both T Tauri stars \citep[e.g. ][]{2005ApJ...628L..65F, 2006ApJ...638..314D}, and brown dwarfs \citep{2010ApJ...720.1668S}, and a substantial sample is available to construct them for Herbig stars as well \citep[e.g.][]{2009A&A...502L..17A,2010ApJ...721..431J}.

We will describe how dust settling is implemented in our radiative transfer code in section \ref{sec:model}, and compare the available approaches that use the turbulent mixing strength. In section \ref{sec:obs}, we will use this model to constrain the turbulent mixing strength from the median SEDs of T Tauri stars, Herbig stars and brown dwarfs, which were constructed in appendix \ref{app:taurus_median} and \ref{app:herbig_median}. We will discuss what the median disk model looks like in section \ref{sec:discussion}, and whether disks around lower mass stars are flatter or more settled than their high-mass counterparts. We summarize our findings in section \ref{sec:conclusion}.

\section{Model}\label{sec:model}

\begin{figure}
  \includegraphics[width=\linewidth]{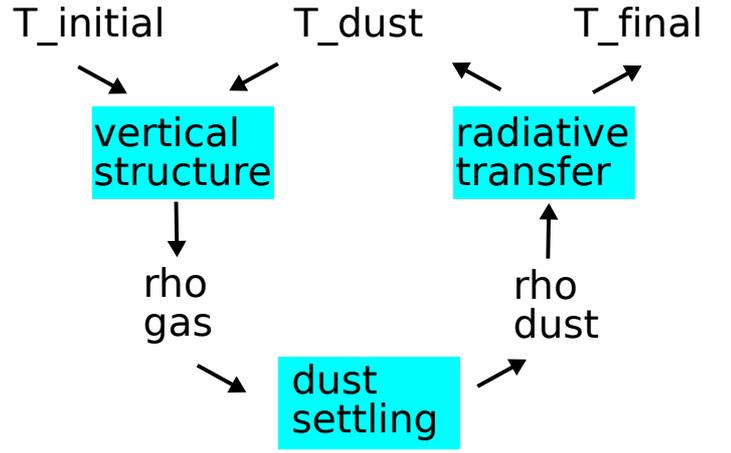}
  \caption[]{Artist impression of vertical structure iterations in MCMax to determine the vertical structure. Model parameters are displayed in black on white, parts of the code as blue blocks. T refers to the temperature, rho to the density.
    \label{fig:iter}}
\end{figure}

\subsection{Disk model}
The disk model used in this paper is MCMax \citep{2009A&A...497..155M}, a 2D Monte Carlo radiative transfer code. It is based on the immediate re-emission procedure from \cite{2001ApJ...554..615B}, combined with the method of continuous absorption by \cite{1999A&A...345..211L}. The code has been benchmarked against other radiative transfer codes for modeling protoplanetary disks \citep{2009A&A...498..967P}. The radial grid around the inner radius and disk wall is refined to sample the optical depth logarithmically. The SED and images are calculated by integrating the formal solution to the equation of radiative transfer by ray-tracing.

      In addition to radiative transfer, the code can explicitly solve for the vertical structure of the disk. With the implicit assumption that the gas temperature is set by the dust temperature, the vertical structure of the gas can be calculated by solving the equation of hydrostatic equilibrium. On top of that, the dust can be settled with respect to the gas, which we will describe in section \ref{sec:mod:set}. After calculating the new structure for the dust, the radiative transfer code can be run to obtain a new dust temperature, and the whole procedure (radiative transfer, gas structure, dust settling) can be iterated until convergence is reached (Figure \ref{fig:iter}). The temperature structure usually converges within five iterations. The exact number of iterations depends on disks parameters. In general, very settled or 'flat' disks take longer to converge because they deviate more from the initial guess.

\subsection{Dust settling}\label{sec:mod:set}
To settle the dust with respect to the gas, we used the formalism by \cite{2004A&A...421.1075D}. We will briefly describe the basic assumptions, and refer to the previously mentioned paper for details. We compare our approach to that of \cite{2010MNRAS.401..143H} in section \ref{sec:compare}, who used the midplane formalism by \cite{1995Icar..114..237D} in combination with radiative transfer models.

\subsubsection{Theory}
The vertical motion of dust particles within the disk is a balance of the downward motion caused by the gravity of the star versus the mainly upward motion caused by turbulent mixing. Comparing the timescales for both processes (the friction time $t_{\rm fric}$ versus the eddy turn-over time $t_{\rm edd}$) results in an expression for the Stokes number St, which describes how well the dust is coupled to the gas at every location in the disk:
\begin{equation}\label{eq:stokes}
{\rm St} = \frac{\rm t_{fric}}{\rm t_{edd}} = 
          \frac{3}{4} \frac{\rm m}{\sigma} \frac{\rm \Omega_k}{\rho_{\rm gas} \rm c_{\rm s}},
\end{equation}
where $m/\sigma$ is the mass-to-surface ratio of the dust grain, $\Omega_{\rm k}= \sqrt{G M_* / r^3}$ the Keplerian frequency, $\rho_{\rm gas}$ is the local gas density and $c_{\rm s} = kT/\mu m_{\rm proton}$ the local sound speed for the mean molecular weight $\mu$. A Stokes number smaller than one means the dust is well coupled to the gas, whereas particles decouple from the gas at larger Stokes numbers. Using this expression for the dust-gas coupling and the standard $\alpha$ prescription from \cite{1973A&A....24..337S} in the form of \cite{1981ARA&A..19..137P} ($\nu=\alpha c_{\rm s} H_{\rm p}$), the diffusion coefficient for a dust particle at every location in the disk can be written as\footnote{We used a Schmidt number of 1+St$^2$ following \cite{2007Icar..192..588Y}, rather than 1+St from \cite{2004A&A...421.1075D}. Because this does not change the point where particles decouple at St=1, it has a very limited influence on our model.}
\begin{equation}\label{eq:diffcoeff}
D = \alpha_{\rm turb} \frac{c_{\rm s} H_{\rm p}}{1+{\rm St}^2},
\end{equation}
where $H_{\rm p}= c_{\rm s} / \Omega_{\rm k}$ is the local pressure scale height and $\alpha_{\rm turb}$ is the turbulent mixing strength. 

Taking the gas density and temperature structure from the radiative transfer and hydrostatic structure calculation, we can now solve for the vertical structure of a dust grain of certain size $\rho_{\rm a}$ at every location in the disk by solving a vertical diffusion equation:
\begin{equation}
\frac{\partial \rho_{\rm a}(z)}{\partial t} = 
\frac{\partial}{\partial z} \left(D_{\rm a}(z) \rho_{\rm gas} 
   \frac{\partial}{\partial z} \left( \frac{\rho_{\rm a}(z)}{\rho_{\rm gas}} \right) \right)
- \frac{\partial}{\partial z} \left(\rho_{\rm a} v_{\rm sett,a}(z) \right),
\end{equation}
where $v_{\rm sett,a}(z)$ is the local settling speed for that dust grain in the absence of turbulent mixing. This equation is then solved in a time-dependent way using implicit integration. Because the timescales on which the disk settles toward equilibrium are shorter than one million years \citep{2004A&A...421.1075D}, we jump toward equilibrium in ten relatively large time steps.

We assumed that the turbulent mixing strength is constant throughout the disk. There is no a priori reason to do so, nor a computational one, and there are indications for variations in both the vertical \citep{2009A&A...496..597F,2011ApJ...743...17S} and radial \citep{2009ApJ...701..260I} direction. However, because we cannot constrain these variations from the data used in this paper, we kept $\alpha$ constant.

\subsubsection{Impact on disk structure}\label{sec:compare}
\begin{figure}
  \includegraphics[width=\linewidth]{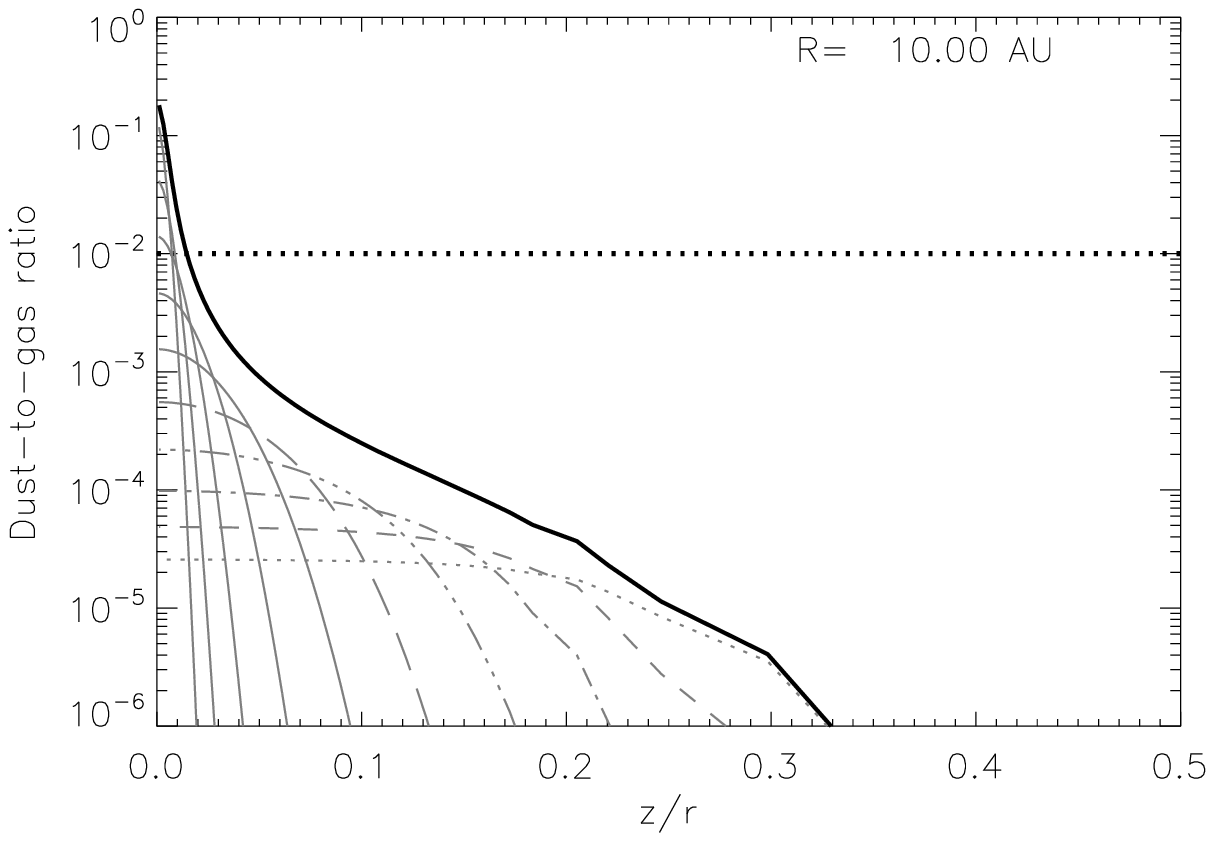}
  \includegraphics[width=\linewidth]{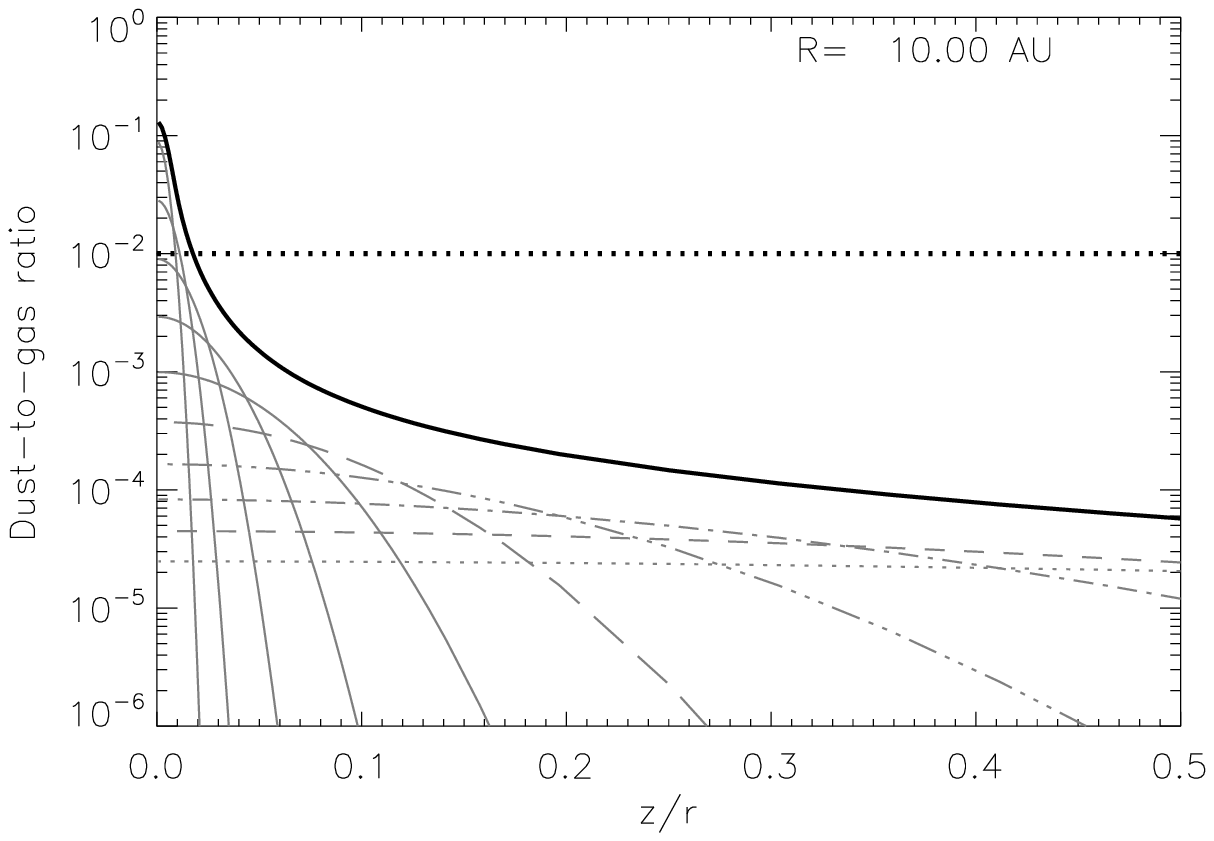}
  \caption[]{Dust-to-gas ratio versus relative height (z/r) in the mid-infrared emitting region of the T Tauri star at 10 AU using the self-consistent settling (top) and the \textit{midplane settling approach} (bottom) with the same gas density distribution. The black solid line represents the total \dtg ratio, the gray lines denote the individual contributions of grains of different sizes, from 0.02 micron (dotted) to 6 mm (solid). The dotted line represents a fully mixed disk, with a dust-to-gas ratio of 0.01. The radial $\tau$=1 surface in the optical is located at a relative height of approximately 0.16 and 0.2.
    \label{fig:settle}}
\end{figure}

The main difference between this approach and that by \cite{1995Icar..114..237D} is that we used the local sound speed and gas density to calculate the dust-gas coupling. Their approach is an analytical solution for modeling the dust distribution in the isothermal midplane, and uses only the midplane density and temperature. Hence, we will refer to the latter as \textit{midplane settling}, though it is sometimes applied outside this regime \citep{2010MNRAS.401..143H}. To compare the different approaches, we have displayed the vertical structure at the first iteration in a vertical slab centered at 10 AU in Figure \ref{fig:settle}. Because this model has not been iterated, the vertical structure of the gas is calculated from the optically thin dust temperature, and corresponds to a Gaussian.

The well-mixed model has a uniform dust-to-gas ratio of 0.01, whereas the settled models show a clear trend from an increased dust-to-gas ratio near the midplane to a decreased ratio near and above the surface, which is located around z/r$\sim$ 0.18 (radial $\tau=1$ surface in the optical). This is caused by stratification, the settling of different dust species to different heights in the disk. In the midplane, the \textit{midplane settling approach} agrees very well with the self-consistent settling, which is also the region that the former has been constructed for \citep{1995Icar..114..237D}. 

Closer to the disk surface, the two settled models start to deviate. With increasing height above the midplane, the gas density decreases. For the self-consistent settling, this means that smaller particles will decouple from the gas. For each particle size, there is a characteristic height above the midplane where the coupling is so weak that they are almost fully depleted \citep{2004A&A...421.1075D}. For particles down to a size of roughly 1 micron, this decoupling takes place below the disk surface, but smaller grains can still be found above the disk surface. 

The \textit{midplane settling approach} shows a different behavior, because the dust-gas coupling is calculated in the midplane only. Particles do not decouple from the gas at a specific height, but keep following a Gaussian distribution all the way through the disk surface. This gives an almost constant dust-to-gas ratio for the smallest particles ($<$ 1 \um), and a much weaker depletion for the intermediate sizes.

\begin{figure}
  \includegraphics[width=\linewidth]{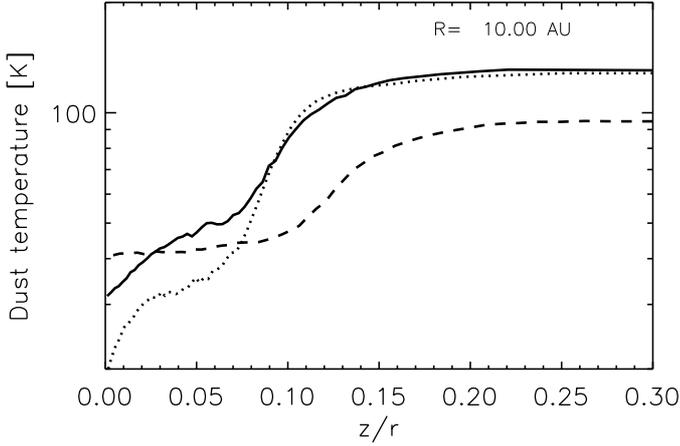}
  \caption[]{Dust temperatures versus relative height in a vertical slab centered at 10 AU for a T Tauri star, showing that stratification in dust size leads to a stratification in dust temperature. Shown are the self-consistent model (solid line), the settling recipe from Dubrulle (dotted line) and a well-mixed model without settling for comparison (dashed line).
    \label{fig:Tmid}}
\end{figure}

Settling the dust also impacts the temperature structure of the disk (Fig. \ref{fig:Tmid}), and therefore its density structure. Non-settled disk models have a more or less constant temperature distribution below the disk surface (Fig. \ref{fig:Tmid}, dash-dotted line). Big grains, however, are efficient coolers, and settling them leads to a colder midplane and warmer surface. The stratification in dust sizes also leads to a gradual increase in temperature from the midplane to the disk surface \citep[see also][]{2010MNRAS.401..143H}. Because the temperature in and above the midplane is no longer isothermal, it becomes important to solve for the vertical structure of the disk using the local scale height, which increases with height. Using only the midplane temperature to calculate the vertical structure of the disk underestimates the scale height within the midplane by up to a factor of two (Fig. \ref{fig:Tmid}). This leads to flatter disks with - in turn - even lower midplane temperatures (Fig. \ref{fig:Tmid}, dotted line). 

\subsection{Dust grain properties}
\begin{figure}
  \includegraphics[width=\linewidth]{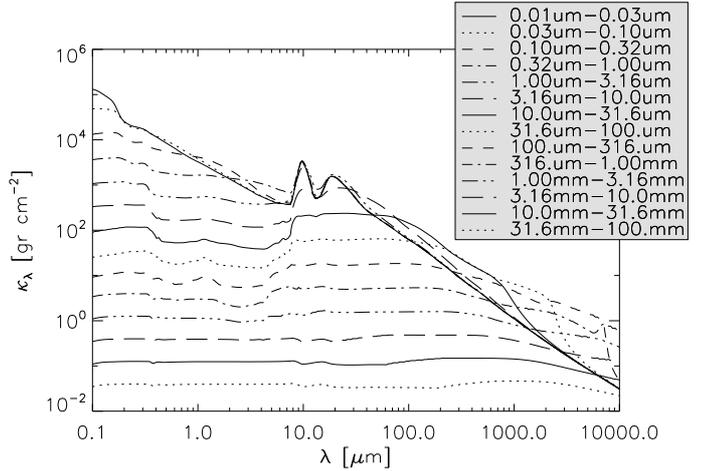}
  \caption[]{Dust opacities for the different grain sizes used in this paper.
  \label{fig:opac}}
\end{figure}

Dust grains in our model are characterized by three different parameters: composition, size and shape. Different grain sizes settle to different heights in the disk (See Fig. \ref{fig:settle}), and their opacities are displayed in figure \ref{fig:opac}.

\subsubsection{Composition and shape}
Our dust grains consist mainly of amorphous silicates, measured for the composition of interstellar dust toward the Galactic center by \cite{2007A&A...462..667M}, co-added with with a continuum opacity source, for which we used amorphous carbon. The ratio of amorphous carbon / amorphous silicates influences the peak strength of the 10-micron silicate feature. We used a mass fraction of 15 \% carbon, a value found from modeling this ratio in Herbig Ae/Be stars \citep{2005prpl.conf.8176M,2009A&A...496..741M}. 
The final composition, with reference to the optical constants, is 11.73 \% MgFeSiO$_{4}$ \citep{1995A&A...300..503D}, 32.6 \% Mg$_{2}$SiO$_{4}$ \citep{1996A&A...311..291H}, 36.5\% MgSiO$_{3}$ \citep{1995A&A...300..503D}, 1.5 \% NaAlSi$_{2}$O$_{6}$ \citep{1998A&A...333..188M}, 15\% C \citep{1993A&A...279..577P}.
The shape of our particles is irregular, and approximated using a distribution of hollow spheres \citep[DHS, ][]{2005A&A...432..909M}. We used a vacuum fraction of 0.7, though this choice does not influence our results because we do not focus on detailed feature shapes.

\subsubsection{Size}
Observations of protoplanetary disks show evidence for (sub)micron and millimeter-sized or larger grains. Intermediate sized grains are not observed, but are predicted by dust coagulation models \citep[e.g. ][]{1984Icar...60..553W,2010A&A...513A..79B}. We therefore used a grain size distribution ranging from sub-micron to millimeter sizes that follows a power law, $f(a) = a^{-q}$, and used a default index of $q=3.5$ similar to the Mathis, Rumpl and Nordsieck (MRN) distribution \citep{1977ApJ...217..425M}. To allow settling of different grain size to a different height in the disk, we divided our size distribution into two bins per order of magnitude. We define the grain radius for the settling calculation with the logarithmic mean of the minimum and maximum grain sizes to  (i.e. $a=1.78$ micron for grain size bin between 1.00 and 3.16 micron). The opacities of these grains were calculated using 10 sizes per bin to avoid strong resonant features that could result from using a single grain size.

\subsection{Disk parameters}
Because our disks are in vertical hydrostatic equilibrium, the only free fitting parameter to control the vertical structure for a given mass is the turbulent mixing strength $\alpha$. The radial distribution of dust and gas is parametrized by a power-law as $\Sigma(r)=r^{-p}$ with the commonly used $p=1$ for a steady-state accretion disk. The inner radius is placed at a dust evaporation temperature of 1500 K, whereas the outer radius is fixed to 300 AU in accordance with \cite{2006ApJ...638..314D}. The total disk mass is a fixed fraction of 1 percent of the stellar mass \citep[e.g.][]{2006ApJ...645.1498S,2011ARA&A..49...67W}, assuming a dust-to-gas ratio of 1:100.

\begin{table}
  \centering
  \begin{tabular}{lll}
    \hline \hline
    Parameter  & Value & Range \\
    \hline
    p                   & 1.0 & 0...2.0 \\
    M$_{\rm disk}$        & 0.01 M$_{\rm star}$ & 0.001...0.1\\
    dust-to-gas          & 0.01 & 0.01...1 \\
    T$_{\rm evap}$        & 1500 & 1000...2500 \\
    R$_{\rm out}$ [AU]    & 300 & 50...1000 \\
    \hline
    a$_{\rm min} [\um]$ & 0.01 & 0.01...10\\
    a$_{\rm max} [\um]$ & 1000  & 10 ... \ten{5} \\
    q                   & -3.5 & -2.5 ... -4.5\\
    \hline
    \alfaturb  & 0.01 & $10^{-6}$ ... 1.0\\
    \hline \hline
  \end{tabular}
  \caption{Parameters of our disk model, and ranges explored. \label{tab:model_parameters}}
\end{table}

\section{Fits to median SEDs}\label{sec:obs}
To compare our disk model to the three different data sets, we used median SEDs. Instead of fitting all individual sources or comparing them to a model grid, we only fitted our model to one median SED to extract 'typical' fit parameters. \cite{2005ApJ...628L..65F,2006ApJS..165..568F} have shown that this approach works well, taking into account the observed spread in mid-infrared colors.

When fitting the data, we tried to keep as many parameters fixed as possible, see Table \ref{tab:model_parameters}. Parameters that could not be fixed because of the different stellar properties are given in Table \ref{tab:stellar_parameters}. These are the inner radius, which we kept at 1500 Kelvin and the dust mass, which is a fixed fraction of the stellar mass of \ten{-4} M$_*$.

\begin{table}
  \centering
  \begin{tabular}{llll}
    \hline \hline
    Parameter  & Brown Dwarf & T Tauri & Herbig Ae \\
    \hline
    T$_{\rm eff}$ [K]        & 3000  & 4000  &  8500 \\
    L$_*$ [L$_{\rm \odot}$]  & 0.025 & 0.9   &   21   \\
    M$_*$ [M$_{\rm \odot}$]  & 0.08  & 0.5   &   2.0  \\
    \hline
    R$_{\rm in}$ [AU]       & 0.011  & 0.07  & 0.30 \\
    M$_{\rm dust}$ [M$_{\rm \odot}$] &  8\e{-6} & 5\e{-5} & 2 \e{-4} \\
    \hline
    distance [pc] & 165 & 140 & 140 \\ 
    M$_{\rm accr}$ [M$_{\rm \odot}$/yr] & 0 & 3\e{-8} & 0 \\
    $\tau_{\rm halo}$ & 0 & 0 & 0.14 \\
    \hline \hline    
  \end{tabular}
  \caption{Stellar-dependent model parameters.
    \label{tab:stellar_parameters}}
\end{table}

\subsection{T Tauri stars}\label{sec:tts}
The Taurus median SED has been fitted using a hydrostatic disk model by \cite{2006ApJ...638..314D}, which demonstrated the need for grain growth and dust settling. Our goal is to reproduce these results, but now in the context of a more self-consistent settling approach to constrain the turbulent mixing strength, and not an arbitrary height for the midplane layer of big grains. We have extended the existing data set from \cite{2006ApJS..165..568F} with sub-millimeter fluxes at 850 \um using the catalog from \cite{2005ApJ...631.1134A}. The data from \cite{1999ApJ...527..893D} include some non-detections and upper limits (see appendix \ref{app:taurus_median}). The updated millimeter median is well fitted by a dust mass of 5\e{-5} M$_{\rm \odot}$, consistent with the result from \cite{2005ApJ...631.1134A}.

\begin{figure}
  \includegraphics[width=\linewidth]{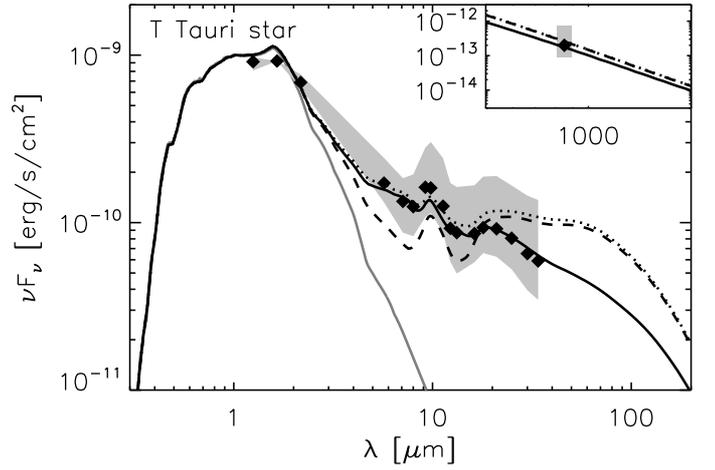}
  \caption[]{Observed median SED of T Tauri stars in Taurus (diamonds) with quartiles (gray area). Overplotted is our best-fit disk model with a turbulent mixing strength of $\alpha=10^{-4}$ and an accretion rate of 3\e{-8} M$_{\rm \odot}$/yr (solid line). Also plotted are a model with $\alpha=10^{-2}$ (dotted line) and without accretion (dashed line). The gray line denotes the stellar photosphere. The inset shows the millimeter regime.
    \label{fig:taurus_median}}
\end{figure}

To fit the median SED in the near-infrared, we needed to include viscous heating (Fig. \ref{fig:taurus_median}). We did not self-consistently solve the radial structure of the disk, but calculated the viscous heating with a mass accretion rate of 3\e{-8} M$_{\rm \odot}$/yr, the same as \cite{2006ApJS..165..568F}. To fit the mid-infrared flux, we flattened the disk by settling the dust. A turbulent mixing strength of $\alpha=10^{-2}$ - a value commonly used for viscous accretion disks - overpredicts the mid-infrared fluxes longward of 20 micron, which arises from the region around $\sim$10 AU. We need to lower the turbulent mixing strength to achieve a good fit, which results in  $\alpha=10^{-4}$. 

Because the 20 micron flux mainly probes the height of the disk surface, there are also other ways to lower it, see also section \ref{sec:alpha_1d-2}. We can reduce the number of small grains, which will lower the optical depth in the upper layers and decrease the visible disk surface. Decreasing the total disk mass would be inconsistent with the millimeter flux, but we can modify the grain size distribution because it is dominated by big grains. Decreasing the power-law index of the grain size distribution removes small grains from the disk surface, but does not significantly affect the mass in big grains near the disk midplane. Changing the power-law index to 3.25 is sufficient to fit the median SED with a turbulent mixing strength of $\alpha=10^{-2}$.

\subsection{Herbig stars}\label{sec:herbigs}
\begin{figure}
  \includegraphics[width=\linewidth]{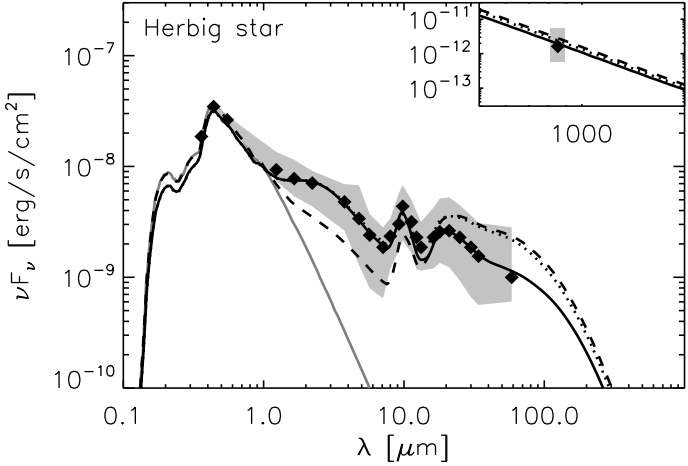}
  \caption[]{Observed median SED of Herbig stars (diamonds) with quartiles (gray area). Overplotted is our best-fit disk model with a turbulent mixing strength of $\alpha=10^{-4}$ and a dust shell with optical depth $\tau=0.14$ (solid line). Also plotted are a model with $\alpha=10^{-2}$ (dotted line) and without a dust shell (dashed line). The gray line denotes the stellar photosphere. The inset shows the millimeter regime.
    \label{fig:herbig_median}}
\end{figure}

Spectral energy distributions for a large number of Herbig stars are available, but a median SED has not been constructed before. We constructed one in appendix \ref{app:herbig_median}, which we present in figure \ref{fig:herbig_median}. To test the assumption that more massive stars have less settled disks, we retained as many disks parameters from the T Tauri stars as possible. We changed (see also table \ref{tab:stellar_parameters}) the disk mass - which scales with the stellar mass and is increased with a factor 5 with respect to the T Tauri star, consistent with the submillimeter data - and the inner radius - which also increases to be consistent with dust evaporation. 

Although accretion does not play a role in the SEDs of Herbig stars because of the increased stellar luminosity and larger inner radius, the near-infrared fluxes (2-8 micron) of hydrostatic disk models underpredict the observations \citep[][see also Fig. \ref{fig:herbig_median}]{2006ApJ...636..348V}. Different mechanisms have been proposed to explain the difference, from an increased rim scale height \citep{2010A&A...516A..48V,2009A&A...502L..17A} to halos \citep{2006ApJ...636..348V,2011A&A...528A..91V} to material within the dust evaporation radius \citep[e.g.][]{2008ApJ...676..490K, 2008ApJ...689..513T}. Although an artificially increased scale height works well for the mid-infrared SED \citep{2009A&A...502L..17A}, it does not work well when fitting simultaneously the near-infrared region of the Herbig median SED (see appendix \ref{app:inner_rim}). Because an inner gaseous disk is beyond the current capabilities of our code, we modeled the near-infrared flux deficit of the hydrostatic model with an optically thin spherical halo. \textit{We emphasize that this halo is a parametrization of optically thin emission from the inner regions, including dust or gas within the inner rim, and refer the interested reader to the appendix}.

The optically thin halo plays the role of accretion in T Tauri stars: it provides near-infrared flux without affecting the flux longward of 20 micron, in contrast to the puffed-up inner rim. We used a small halo of $0.3...1$ micron grains between 0.14 and 0.3 AU\footnote{Because the halo spans a very narrow temperature range, its near-infrared spectrum is virtually indistinguishable from that of a puffed-up inner rim (see Fig. \ref{fig:median_shadow}).} that has an optical depth of 0.14 in the visual and a dust mass of 4\e{-12} M$_{\rm \odot}$(Figure \ref{fig:herbig_median}). As a result of the fitting procedure, we found the same turbulent mixing strength for the disk of $\alpha=10^{-4}$ as for the T Tauri stars. A higher mixing strength of $\alpha=10^{-2}$ overpredicts the Spitzer IRS fluxes from 20 to 40 micron and the IRAS flux at 60 micron.

\subsection{Brown dwarfs}
Disks around brown dwarfs are more difficult to detect owing to their low luminosity. Complete samples down to photospheric values only exist up to 8 micron, and are incomplete at 24 micron \citep[e.g.][]{2006AJ....131.1574L,2007ApJ...660.1517S,2008ApJ...675.1375L}. Furthermore, very few millimeter detections exist \citep{2003ApJ...593L..57K, 2006ApJ...645.1498S}. We used the median SED constructed by \cite{2010ApJ...720.1668S} for Chameleon I, where the median 24 micron flux should be treated as an upper limit. For convenience, we also plot the millimeter upper limit from \cite{2003ApJ...593L..57K}, scaled to the distance of the Chameleon I star-forming region.

\begin{figure}
  \includegraphics[width=\linewidth]{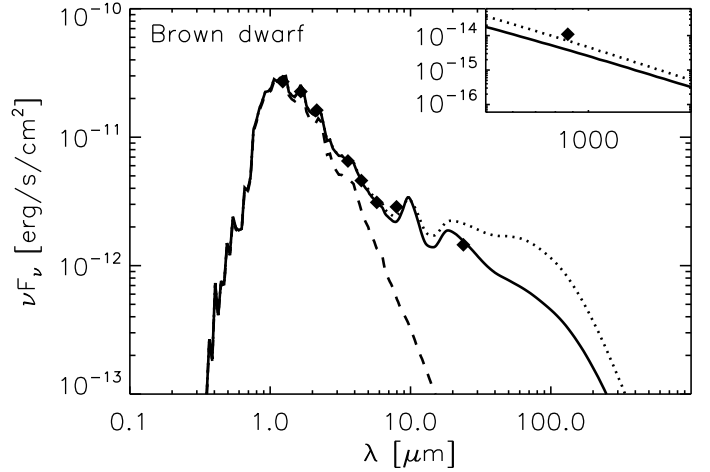}
  \caption[]{Observed median SED of brown dwarfs in Chamaeleon I (diamonds), where the 24 micron and millimeter point should be treated as upper limits. Overplotted is our best-fit disk model with a turbulent mixing strength of $\alpha=10^{-4}$ (solid line). Also plotted for comparison is a model with $\alpha=10^{-2}$ (dotted line). The dashed line denotes the stellar photosphere. The inset shows the millimeter regime.
    \label{fig:BD_median}}
\end{figure}

Unlike the Herbig and T Tauri stars, hydrostatic disk models of brown dwarfs do not show a flux deficit in the near-infrared compared to the observed median. We scaled the disk mass and inner radius to the brown dwarf regime (see Table \ref{tab:stellar_parameters}). The stellar photosphere was fitted by a NextGen stellar atmosphere \citep{1999ApJ...512..377H}, and we adjusted the luminosity down to 0.025 L$_{\rm \odot}$. We fitted the median shortward of 8 micron without any additional adjustments (see figure \ref{fig:BD_median}). The assumed dust mass of 8\e{-6} M$_{\rm \odot}$ is consistent with the upper limit from \cite{2003ApJ...593L..57K}, and amounts to only 2-3 earth masses of solid material.

We found that a turbulent mixing strength of $\alpha=10^{-4}$ is consistent with the 24 micron upper limit, whereas a mixing strength of $\alpha=10^{-2}$ clearly overpredicts this upper limit. We found the same turbulent mixing strength as for the earlier spectral types.

\section{Discussion}\label{sec:discussion}
Although we found the same turbulent mixing strength for the three different samples, this does not necessarily mean that these disks are equally flat. The vertical structure is a product of more than just the turbulent mixing strength, most notably the local density, temperature and Keplerian frequency. In the following section we will describe the disk properties, and show that disks around lower mass stars appear more settled at a specific radius, consistent with \cite{2010ApJ...720.1668S}, but that the structure in the mid-infrared emitting region is remarkably self-similar.

We will also discuss why our choice of grain size distribution does not affect our conclusion that the turbulent mixing strength does not vary along the stellar mass range in section \ref{sed:discuss_gsd} and its implication for the planet-forming potential of the disk in section \ref{sec:planets}. The dependence of the fitted turbulent mixing strength on other disk parameters is discussed in section \ref{sec:alpha_1d-2}.

\subsection{Disk properties at fixed radius}
\begin{figure*}
  \includegraphics[width=0.33\textwidth]{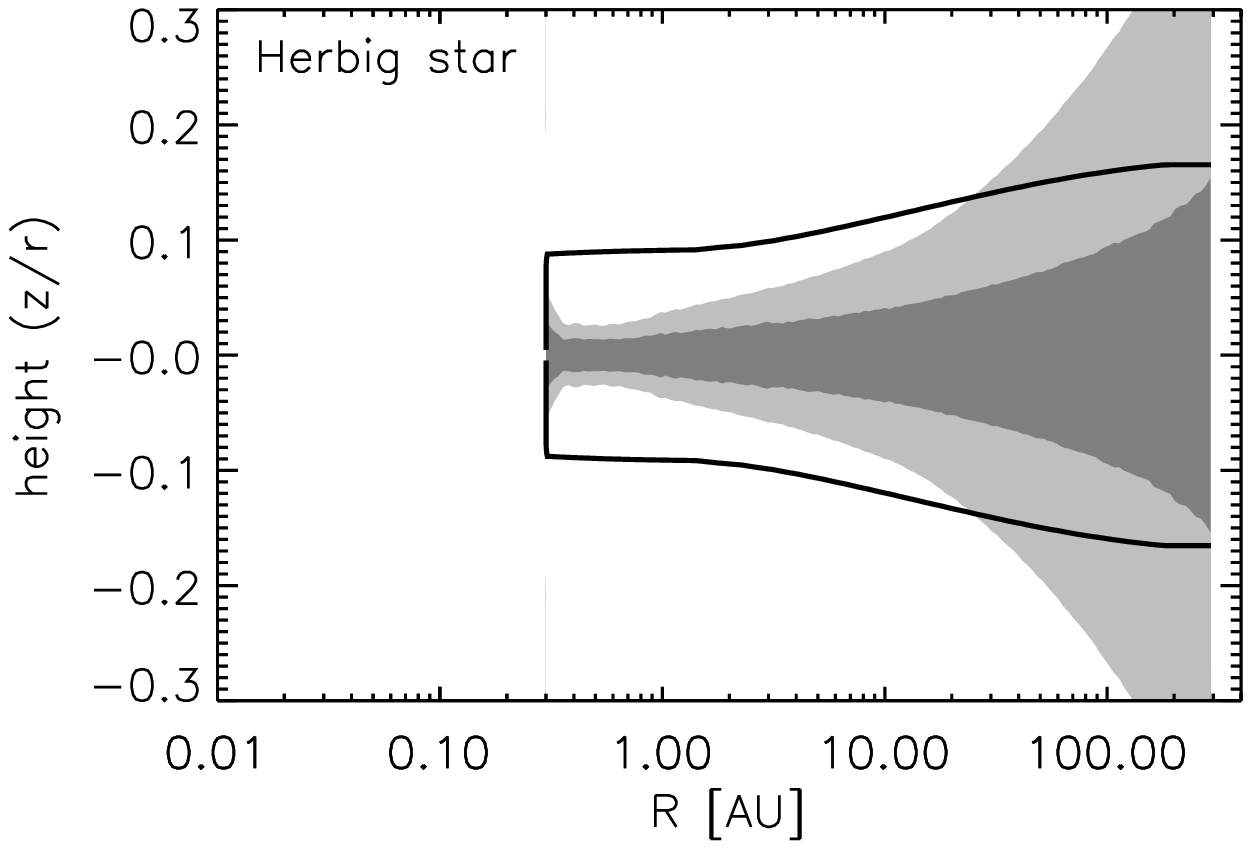}
  \includegraphics[width=0.33\textwidth]{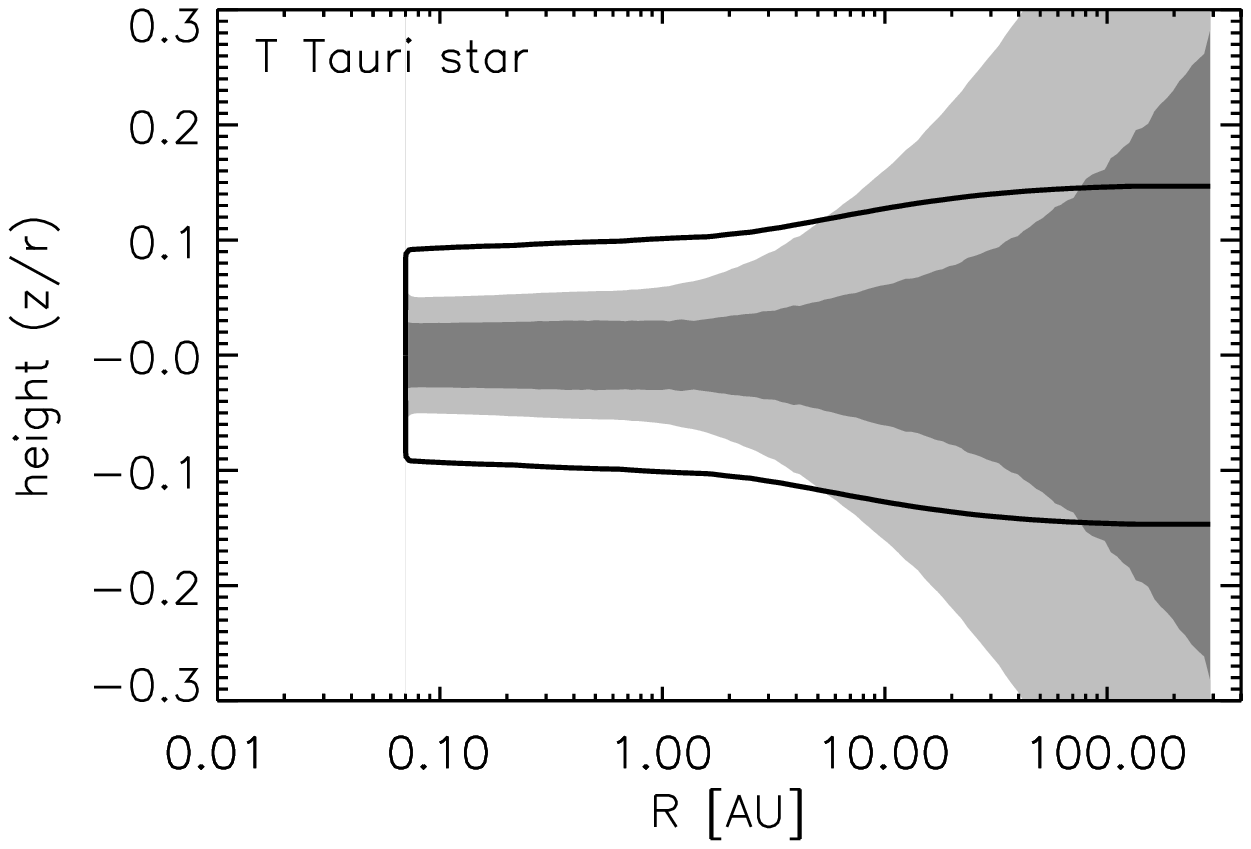}
  \includegraphics[width=0.33\textwidth]{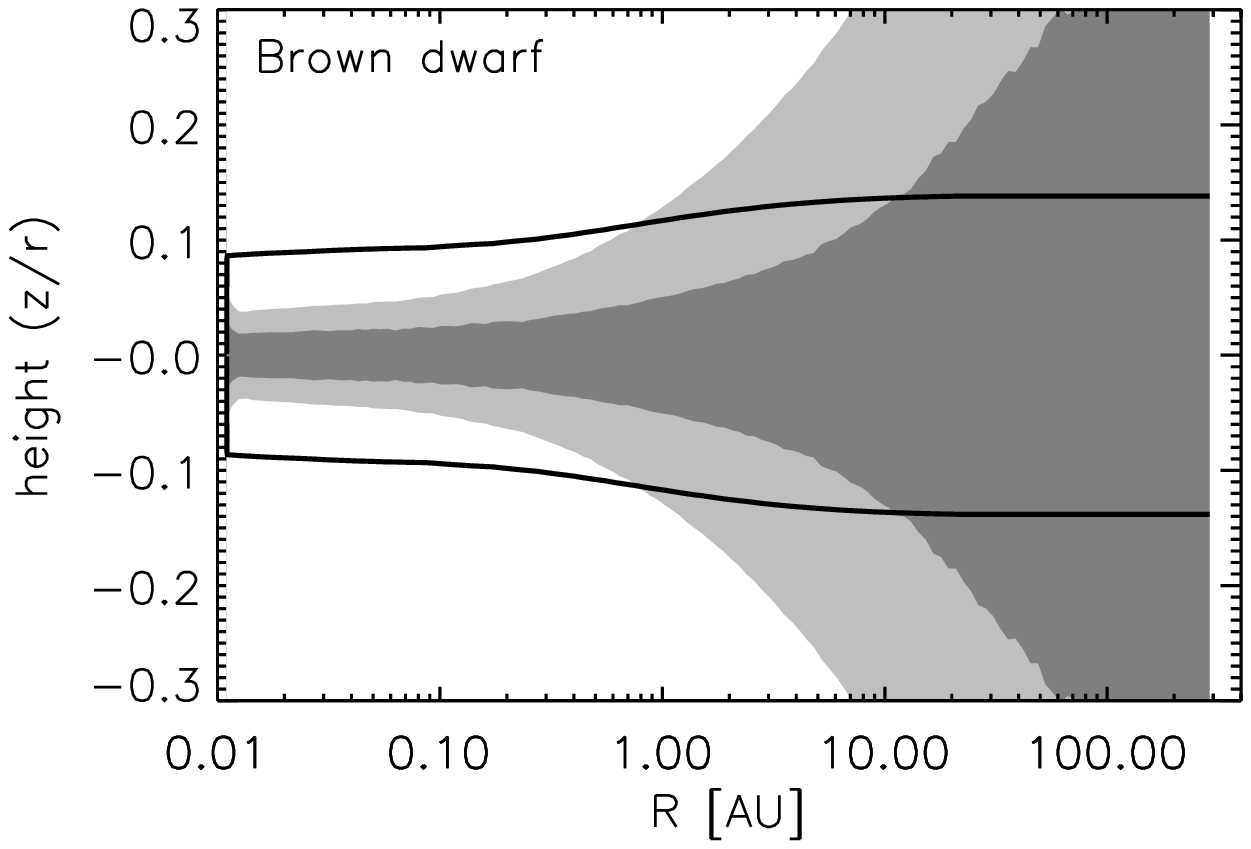}
  \caption[]{Location of the dust and gas in the disk of the best-fit model of a T Tauri star (center), Herbig star (left) and a brown dwarf (right). Displayed in gray is the gas located within one (dark) and two (light) pressure scale heights.  The solid lines marks the location of the radial $\tau=1$ surface in the visual.
    \label{fig:dustgas}}
\end{figure*}

To compare the dust and gas distribution in our best-fit models, we display their pressure scale height and the radial $\tau=1$ surfaces in figure \ref{fig:dustgas}. The pressure scale height is highest in the brown dwarf disk and lowest in the Herbig star, reflecting the differences in stellar mass. Although temperatures are also lower for less massive stars, this is not enough to compensate the lower gravitational potential in which the disk resides, and the gas remains more flaring. 

\begin{figure*}
\begin{minipage}{\textwidth}
  \centering

  \includegraphics[width=0.4\textwidth]{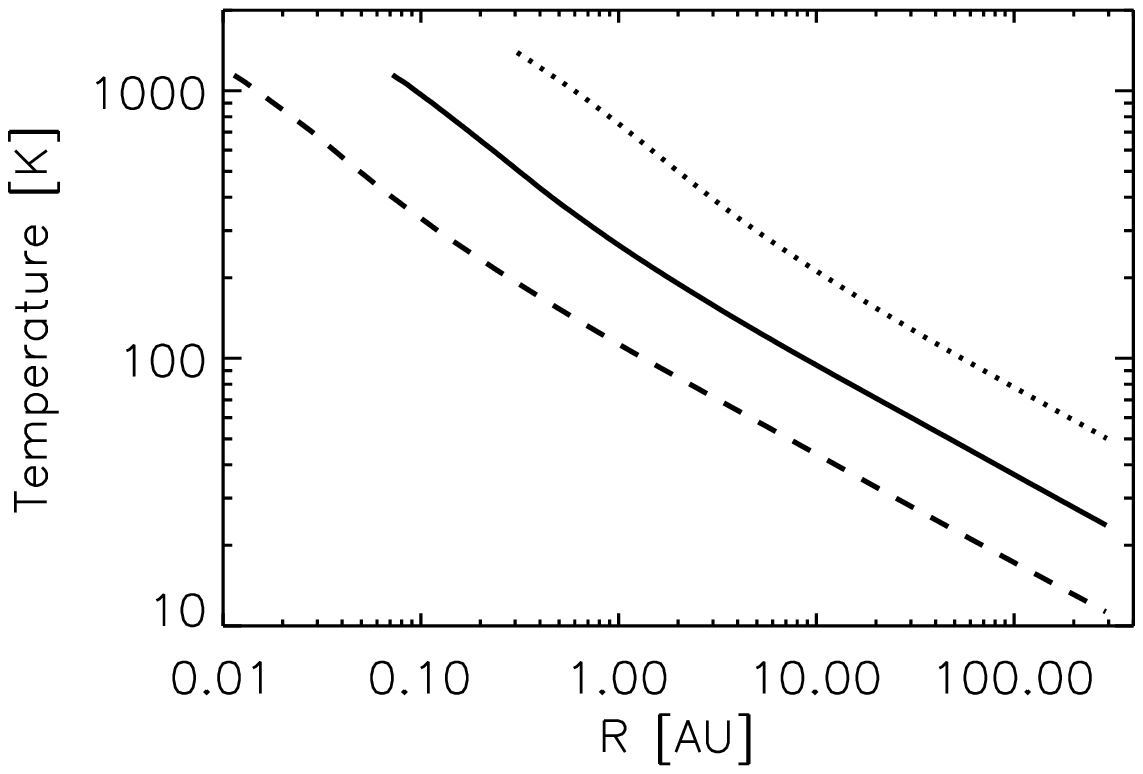}
  \hspace{1cm}
  \includegraphics[width=0.4\textwidth]{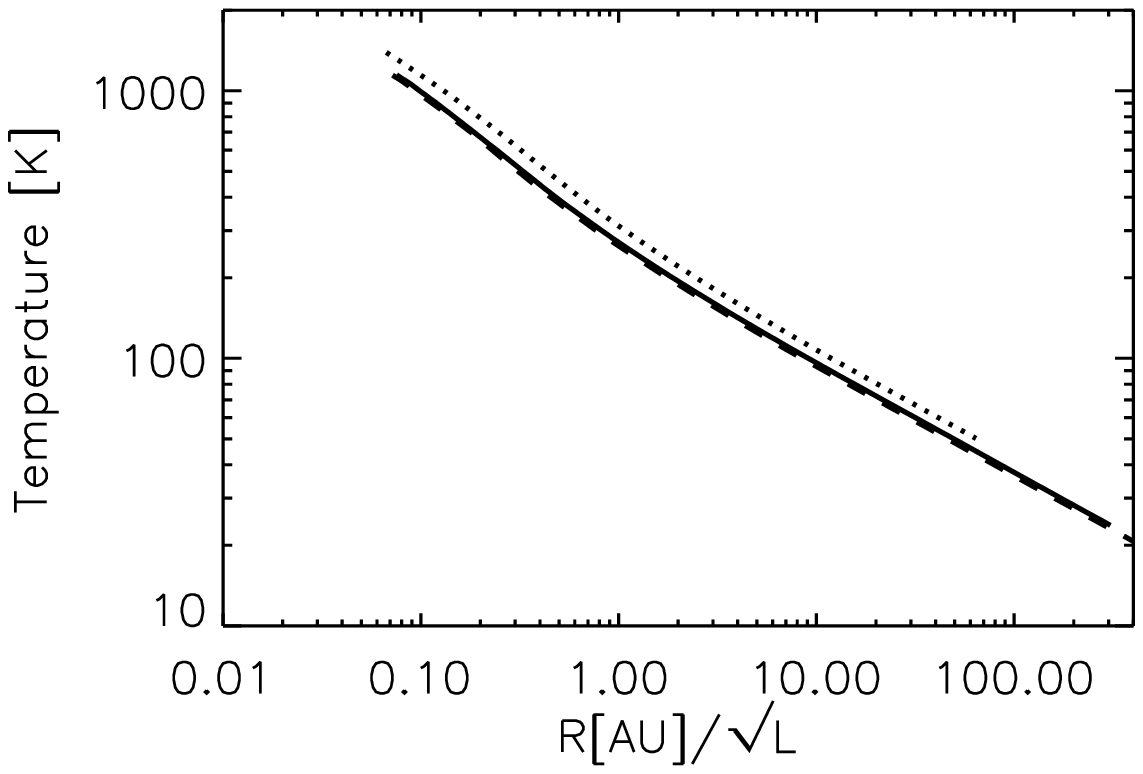}
  \put(-460,125){a)}
  \put(-215,125){b)}
  \caption[]{Radial temperature profile of the optically thin dust for the best-fit model of a T Tauri star (solid line), Herbig stars (dotted line) and a brown dwarf (dash-dotted line). The left panel shows the real temperature, the right panel shows the self-similar solutions. 
    \label{fig:comp_temp} }

  \includegraphics[width=0.4\textwidth]{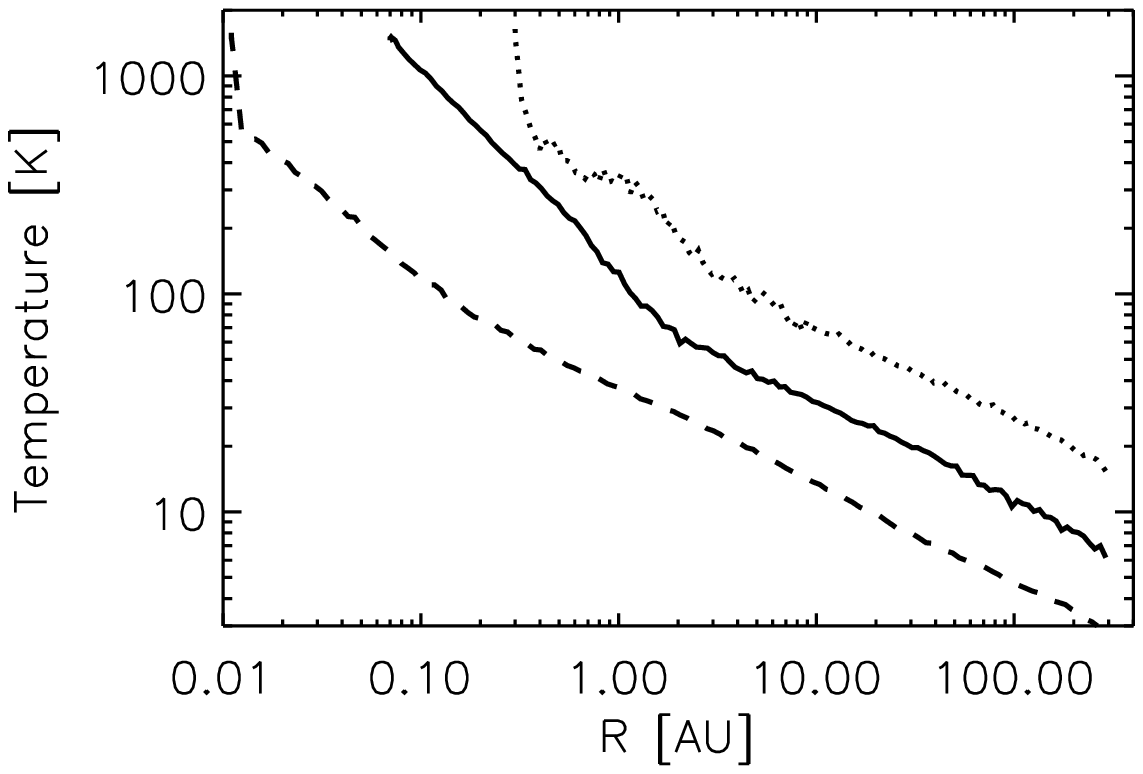}
  \hspace{1cm}
  \includegraphics[width=0.4\textwidth]{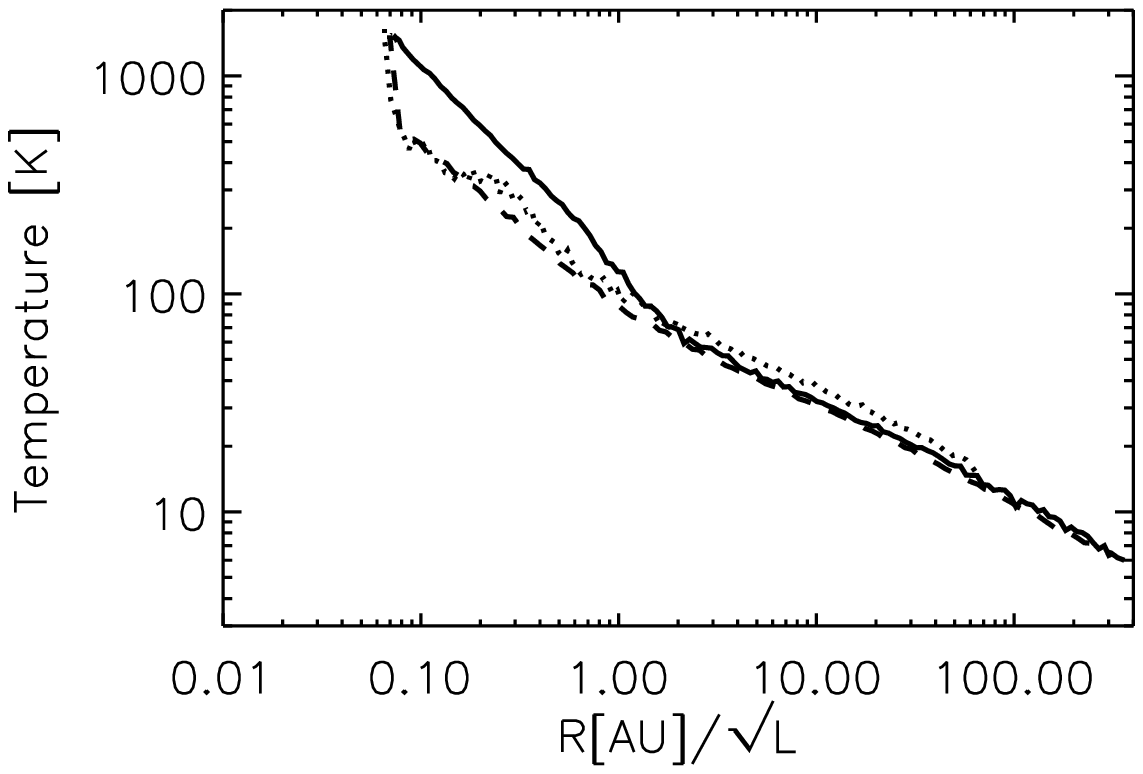}
  \put(-460,125){a)}
  \put(-215,125){b)}
  \caption[]{Radial temperature profile in the midplane for the best-fit model of a T Tauri star (solid line), Herbig stars (dotted line) and a brown dwarf (dash-dotted line). The left panel shows the real temperature, the right panel shows the self-similar solutions. 
    \label{fig:comp_tmid} }

  \includegraphics[width=0.4\textwidth]{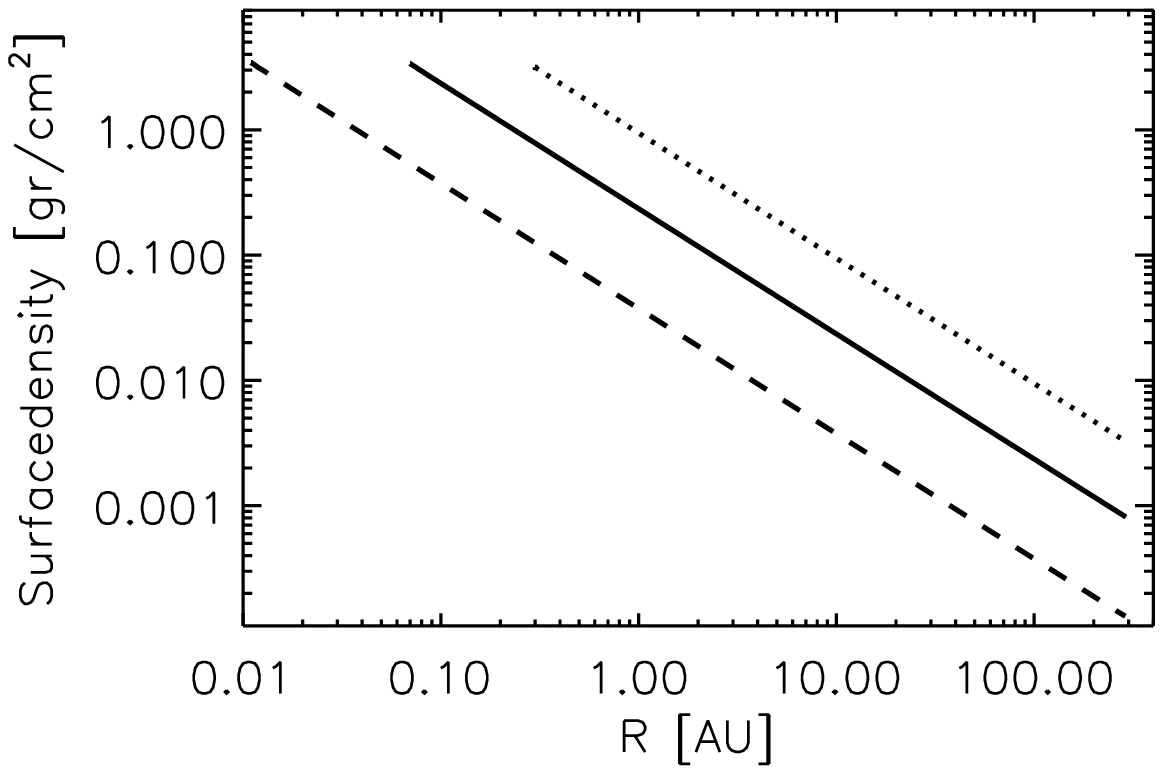}
  \hspace{1cm}
  \includegraphics[width=0.4\textwidth]{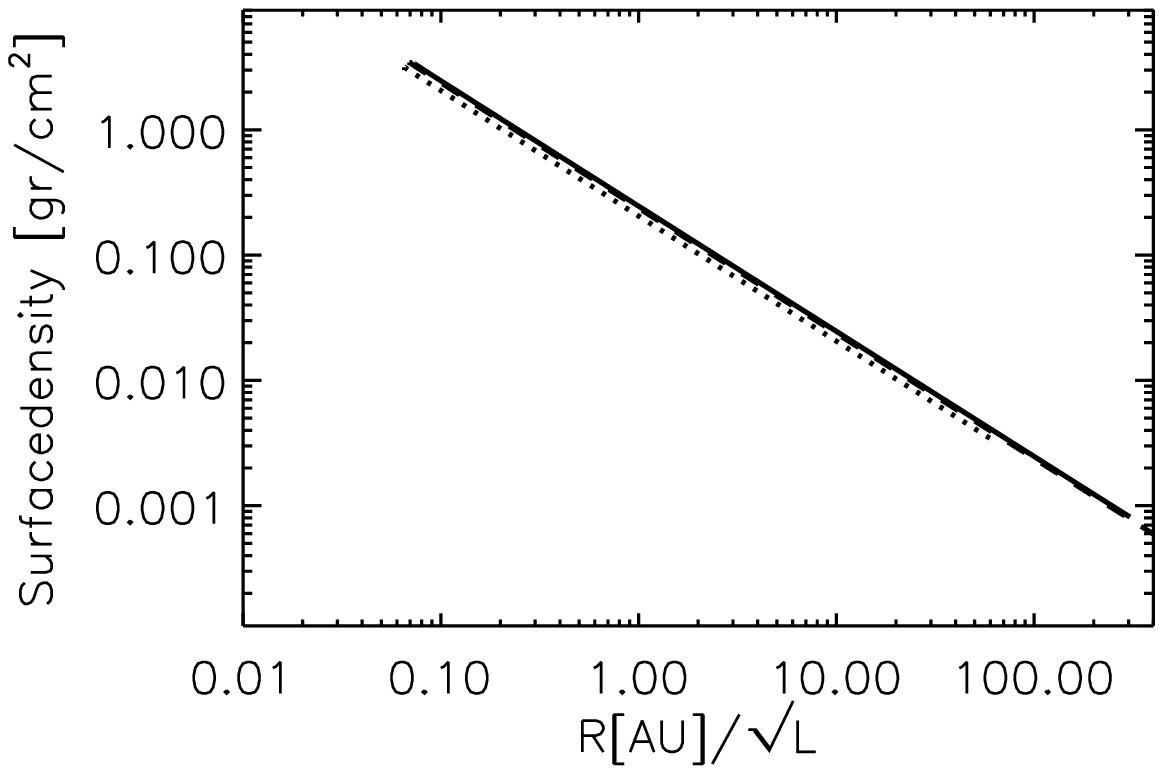}
  \put(-460,125){a)}
  \put(-215,125){b)}
  \caption[]{Radial surface density profile for the best-fit model of a T Tauri star (solid line), Herbig stars (dotted line) and a brown dwarf (dash-dotted line). The left panel shows the real surface density, the right panel shows the self-similar solutions. 
    \label{fig:comp_sigma} }

  \includegraphics[width=0.4\textwidth]{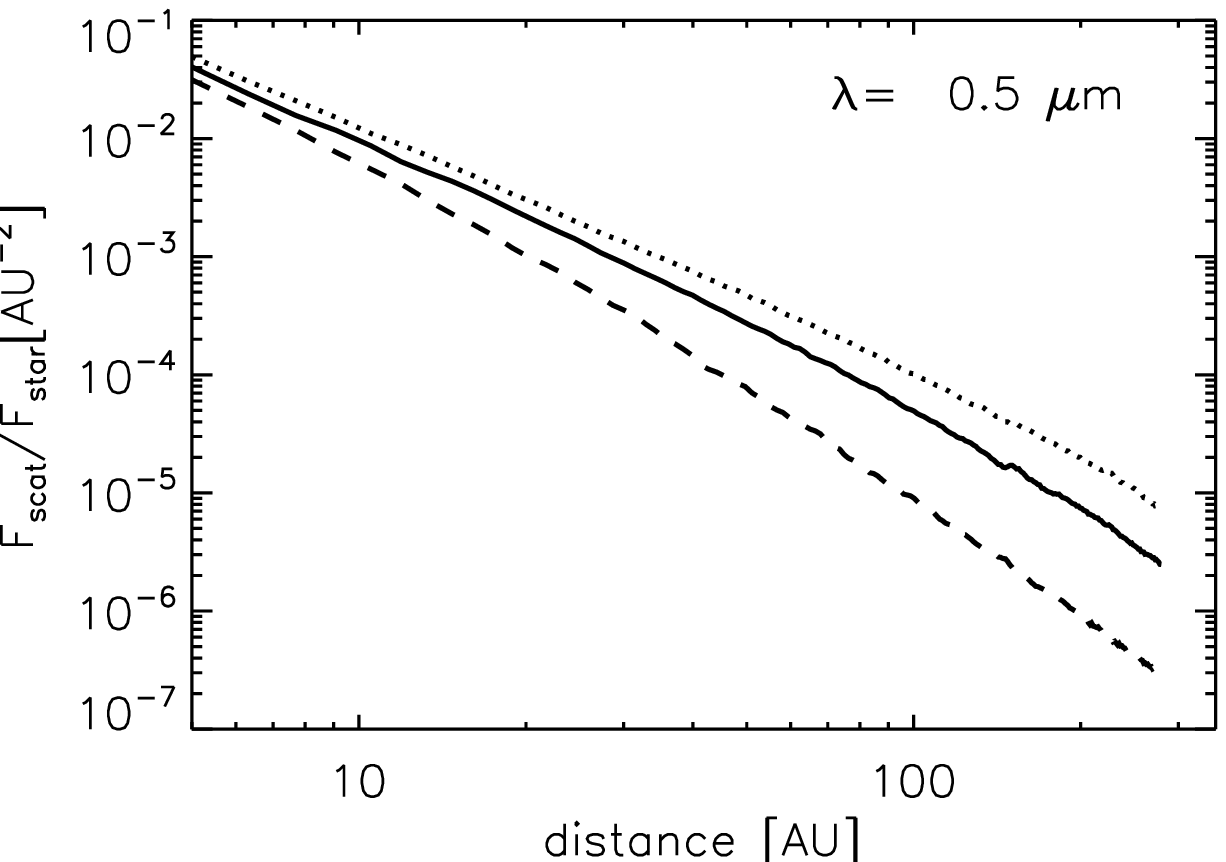}
  \hspace{1cm}
  \includegraphics[width=0.4\textwidth]{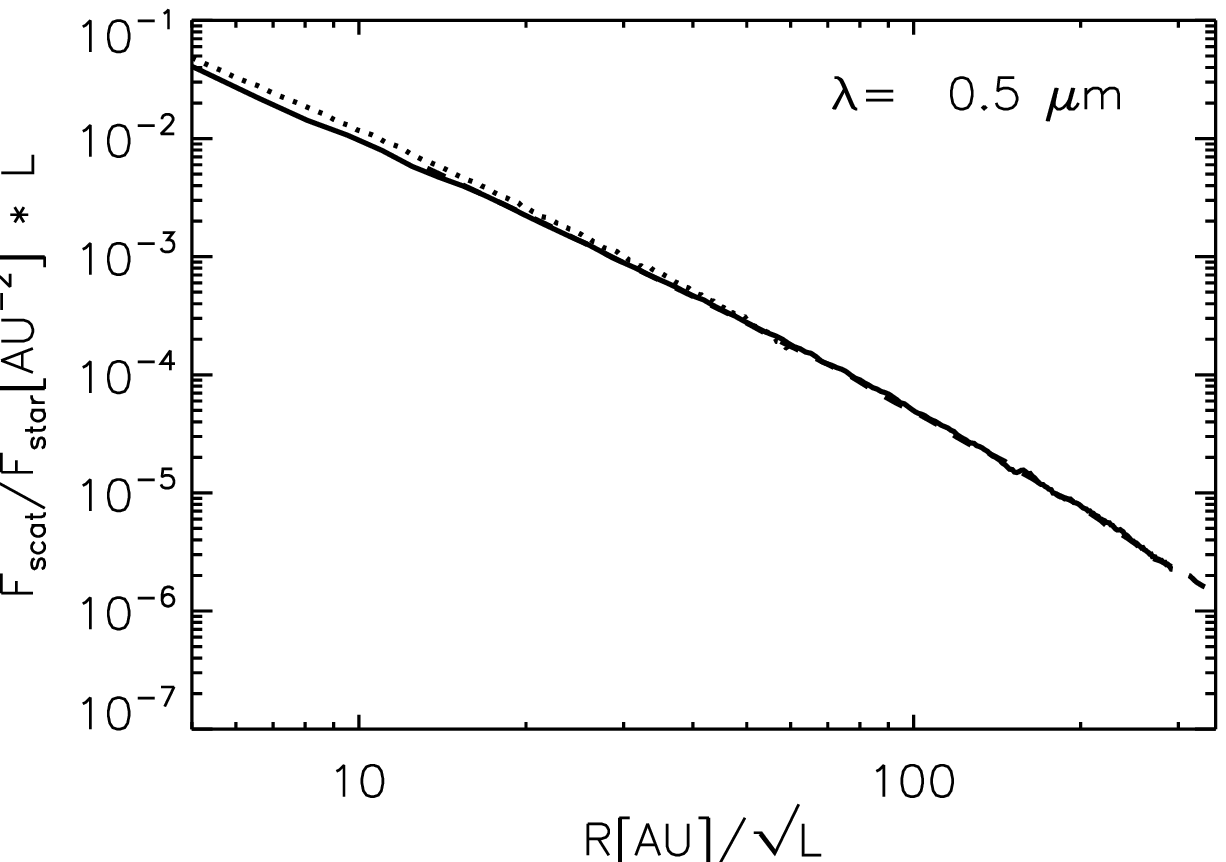}
  \put(-460,125){a)}
  \put(-215,125){b)}
  \caption[]{Radial intensity profile in scattered light at 0.5 micron for the best-fit model of a T Tauri star (solid line), Herbig stars (dotted line) and a brown dwarf (dashed line). The left panel shows the real intensity as fraction of the stellar flux at that wavelength F$_{\rm scat}$(r)/F$_*$ versus r), the right panel shows the self-similar solutions (F$_{\rm scat}$(r)/F$_* \cdot (L_*/L_{\odot})$ versus r/$\sqrt{L_*/L_{\odot}}$
    \label{fig:comp_scat} }
\end{minipage}
\end{figure*}

The dust, on the other hand, follows a different path: the $\tau=1$ surface in the outer disk reaches a relative height ($z/r$) of 0.17, 0.15 and 0.14 for the Herbig star\footnote{It has to be noted that the halo increases the calculated $\tau=1$ surface for the Herbig star.}, T Tauri star and brown dwarf, respectively. This indicates that disks around less massive stars are a little flatter, though the difference remains within 20\%. Compared to the much higher pressure scale height at the same radius, the disk of lower mass stars therefore appears more settled, because the disk surface reaches down to one pressure scale height for the Herbig star, but to less than half a pressure scale height in the brown dwarf. This can be explained by lower (surface) densities and temperatures (see figure \ref{fig:comp_tmid}a, \ref{fig:comp_sigma}a and Table \ref{tab:disk_properties}) that make the dust decouple closer to the midplane, as well as a lower optical depth.

The shape of the disk surface can also be viewed in scattered-light images. We display the radial profiles of synthetic scattered-light images at 0.5 \um in figure \ref{fig:comp_scat}a. The radial profiles are scaled to the photospheric flux at the same wavelength, such that it traces only the \textit{shape} of the disk surface, and not the properties of the central star. A clear trend is visible at all radii, with higher fluxes and shallower slopes for Herbig stars and lower fluxes and steeper slopes for brown dwarfs (see also Table \ref{tab:disk_properties}). This indicates that the disk surface at a specific location is less flaring.

\subsection{Self-similar solutions in the MIR emitting region}\label{sec:self_similar}
Because the mid-infrared emission probes a different region in all three samples, it is also interesting to compare regions with similar temperatures, which are located at different radii for different stars. These regions can be identified by using that the distance $r$, at which a specific flux from the star is received, scales with the square root of the stellar luminosity $L_*$. By scaling the radius to the value of the stellar flux, $r_{\rm T}=r/\sqrt{L_*/L_{\odot}}$, we can construct dimensionless equivalents of our disk properties (denoted by a \textasciicircum ), and check if they are self-similar across the stellar mass range.

For example, the optically thin dust temperature for the T Tauri star can be described as (see Fig. \ref{fig:comp_temp}a)
\begin{equation}
  T_{\rm dust}(r) = 307~K * (r/\rm AU)^{-0.5},
\end{equation} 
By scaling the radius to the intensity of the radiation field, we can retrieve a self-similar solution for the optically thin dust temperature profile
\begin{equation}
\begin{split}
  T_{\rm dust}(r)
  &= 307~{\rm K} * \left(r_{\rm T}/{\rm AU} * \sqrt{L_*/L_{\rm \odot}}\right)^{-0.5} \\
  &= 315~{\rm K} * (r_{\rm T}/{\rm AU})^{-0.5} = \hat{T}_{\rm dust}(r_{\rm T}),
\end{split}
\end{equation}
where the temperature at 1 AU for the T Tauri star is close to the selfconsistent solution because its luminosity is close to 1 $L_\odot$. As can be seen in figure \ref{fig:comp_temp}b, these scaled solutions agree to within 20\% for the T Tauri star, Herbig star and brown dwarf.

\subsubsection{Self-similar temperature and surface density}
With the same procedure, we can also compare different quantities to see if they are also self-similar in all three samples. We find that also the midplane temperatures for all three samples (see Fig. \ref{fig:comp_tmid}b) can be described with a self-similar function:
\begin{equation}\label{eq:Tmid}
\hat{T}_{\rm mid}(r_{\rm T}) = 90~{\rm K} * (r_{\rm T}/{\rm AU})^{-0.44},
\end{equation}
which is a strong indication that this must be a consequence of a self-similar vertical structure across the stellar mass range, which will we explore below in more detail. Interestingly, this scaling law also works for the surface density (Fig. \ref{fig:comp_sigma}), such that regions with similar temperatures also have similar surface densities:
\begin{equation}\label{eq:sigma}
\hat{\Sigma}(r_{\rm T}) = 0.25~{\rm g/cm}^2 * (r_{\rm T}/{\rm AU})^{-1},
\end{equation}
This is because across the three samples, the luminosity scales roughly with stellar mass squared\footnote{Although the main-sequence mass luminosity relation for solar mass stars follows L $\propto$ M$^{3.5...4}$, the T Tauri stars in our sample are pre-main sequence and do not follow this relation, whereas the brown dwarfs are in a different scaling regime altogether.}. Since the disk mass is a fixed fraction of the stellar mass, the surface density scales as
\begin{equation}\label{eq:sigma}
\hat{\Sigma}_{\rm 1AU} = \left(\frac{M_\odot}{M_*}\right) ~ \Sigma_{\rm 1AU} 
                      = \sqrt{\frac{L_\odot}{L_*}} ~ \Sigma_{\rm 1AU},
\end{equation}
However, the region corresponding to the same temperature lies farther away from the star, compensating for the change in disk mass:
\begin{equation} \begin{split}
\hat{\Sigma}(r_{\rm T}) &= \hat{\Sigma}_{1AU} ~ {r_{\rm T}}^{-1} 
= \left(\frac{\Sigma_{1AU}}{\sqrt{L_*/L_{\odot}}}\right)\left(\frac{r}{\sqrt{L_*/L_{\odot}}}\right)^{-1}\\ 
& = \Sigma_{1AU}~r^{-1} = \Sigma(r),
\end{split} \end{equation}
Thus the self-similar solutions for the surface density agree to within 20 \%.

\subsubsection{Self-similar scale height}
If we now look at a vertical slab in the disk located at $r_{\rm T}=r/\sqrt{L_*/L_{\odot}}$, we find that two of the main parameters for the settling calculation in equations \eqref{eq:stokes} and \eqref{eq:diffcoeff} - the gas surface density \eqref{eq:sigma} and temperature \eqref{eq:Tmid} - are self-similar. The third important parameter, the Keplerian frequency, is not the same, but scales with $\sqrt{\rm L_*/L_{\odot}}$
\begin{equation} \begin{split}
 \hat{\Omega}_{\rm k}(r_{\rm T}) &= \sqrt{ \frac{GM_*}{r_{\rm T}^3}}
= \sqrt{ G \frac{M_\odot}{\sqrt{L_*/L_\odot} } \left( \frac{r}{ \sqrt{L_*/L_\odot} } \right)^{-3} } \\
&= \sqrt{\frac{G M_\odot}{r^3} } \sqrt{L_*/L_{\odot}} 
= \Omega_{\rm k}(r) ~ \sqrt{L_*/L_{\odot}},
\end{split} \end{equation}
which is equivalent to $\rm r/r_{\rm T}$. The gas pressure scale height at this location is therefore not self-similar, but the \textit{relative} scale height $\rm H_p/r$ is:
\begin{equation}
\frac{\hat{H}_{\rm p}(r_{\rm T})}{r_{\rm T}}= 
\frac{\hat{c}_{\rm s}(r_{\rm T})}{r_{\rm T}\hat{\Omega}_{\rm k}(r_{\rm T})} =
\frac{c_{\rm s}(r)}{ r_{\rm T}~ (\Omega_{\rm k}(r)~r/r_{\rm T})} =
\frac{c_{\rm s}(r)}{r\Omega_{\rm k}(r)} =
\frac{H_{\rm p}(r)}{r}.
\end{equation}
This means that if we express the height above the midplane ($z$) in terms of the relative height ($z/r$), the local scale height, and therefore the vertical structure of the gas, is self-similar:
\begin{equation}
\hat{H}_{\rm p}(r_{\rm T},z_{\rm T}/r_{\rm T})=H_{\rm p}(r_{\rm T},z_{\rm T}/r_{\rm T}).
\end{equation} 

\subsubsection{Self-similar dust settling}
\begin{figure}
  \includegraphics[width=\linewidth]{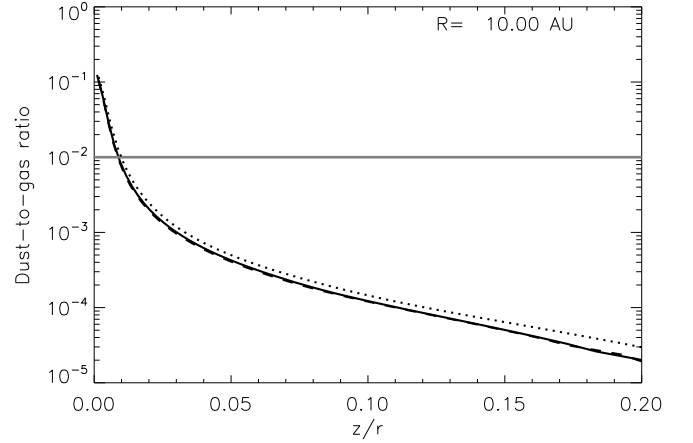}
  \caption[]{Dust-to-gas ratio at $r_{\rm T}=10$ AU for the best-fit model of a T Tauri star (solid line), Herbig star (dotted line) and brown dwarf (dashed line). The gray line denotes a well-mixed model with a dust-to-gas ratio of 0.01. The surface temperature at this location corresponds to roughly 100 K, and therefore emits mainly in the mid-infrared.
    \label{fig:comp_settle}}
\end{figure}

The vertical structure of the dust in the mid-infrared emitting region at $r_{\rm T}=10 \rm AU$ is shown in figure \ref{fig:comp_settle}. Like that of the gas, the vertical structure of the dust also appears self-similar when expressed in terms of its relative height $z/r$. This is because also the Stokes number (Eq. \eqref{eq:stokes})
\begin{equation} \begin{split}
    \hat{\rm St}(r_{\rm T},z_{\rm T}/r_{\rm T})
    &=\frac{3}{4} \frac{\rm m}{\sigma} 
      \frac{\hat{\Omega}_{\rm k}(r_{\rm T})}
           {\hat{\rho}_{\rm gas}(r_{\rm T},z_{\rm T}/r_{\rm T})~c_{\rm s}(r_{\rm T},z_T/r_{\rm T})}\\
    &=\frac{3}{4} \frac{\rm m}{\sigma}
      \frac{\Omega_{\rm k}(r)/\sqrt{L_*/L_{\odot}}}{\rho_{\rm gas}(r,z/r)/\sqrt{L_*/L_{\odot}}~c_{\rm s}(r,z/r)}\\
    &={\rm St}(r,z/r)
\end{split} \end{equation}
is self-similar because the gas density scales with $1/H_{\rm p}$. The same is true for the diffusion coefficient \eqref{eq:diffcoeff}
\begin{equation} \begin{split}
\hat{D}(r_{\rm T},z_{\rm T}/r_{\rm T}) 
&= \alpha_{\rm turb} 
   \frac{\hat{c}_{\rm s}(r_{\rm T},z_{\rm T}/r_{\rm T}) \hat{H}_{\rm p}(r_{\rm T},z_{\rm T}/r_{\rm T})}
        {1+\hat{\rm St}^2(r_{\rm T},z_{\rm T}/r_{\rm T})} \\
&= \alpha_{\rm turb} 
   \frac{c_{\rm s}(r_{\rm T},z_{\rm T}/r_{\rm T}) H_{\rm p}(r_{\rm T},z_{\rm T}/r_{\rm T})}
        {1+{\rm St}^2(r_{\rm T},z_{\rm T}/r_{\rm T})} \\
&= D(r,z/r) 
\end{split} \end{equation}
because it is only a product of self-similar quantities when expressed in units of $r_{\rm T}$ and $z_{\rm T}/r_{\rm T}$. Consequently, the vertical structure of both gas and dust in the mid-infrared emitting regions of protoplanetary disks are self-similar. 

This result is consistent with that of \cite{2011ApJS..195....3F}, who also founnd no variation in the degree of settling between brown dwarfs and T Tauri stars, but partly contradicts \cite{2010ApJ...720.1668S}. These authors also found the same relative scale height ($H/r$) for the dust in the mid-infrared emitting region. However, the required reduction in dust scale height compared to their fully flared models implies a difference in gas scale height between brown dwarfs and T Tauri stars, which we do not find.

The self-similarity of vertical structure is further illustrated by synthetic scattered-light images that trace the shape of the dust disk surface (Fig. \ref{fig:comp_scat}). If we plot the radial intensity profile as a function of the scaled radius $r_{\rm T}$, corrected for the relative height\footnote{The relative height $z/r$ scales with $r_{\rm T}/r= 1/\sqrt{L_*/L_\odot}$, while the intensity in scattered light scales with $(r_{\rm T}/r)^2 = 1/(L_*/L_{\odot})$. } ($\rm F_{\rm scat}/F_* * L_*/L_\odot$), we see that the radial profiles for the three different stars line up nicely (Figure \ref{fig:comp_scat}b), and that the shape of the disk surface is indeed self-similar. 

\begin{figure}
  \includegraphics[width=\linewidth]{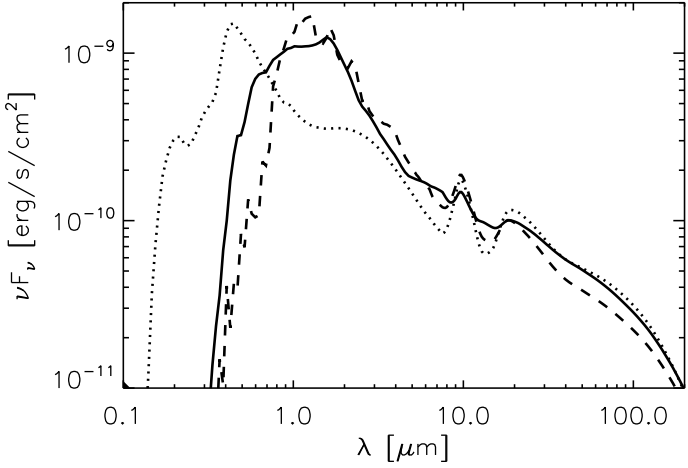}
  \caption[]{Best-fit model SEDs for the Taurus median (solid line), Herbig median (dotted line) and brown dwarf median (dashed line), scaled to a luminosity of 1 $L_\odot$ and a distance of 140 pc.
    \label{fig:comp_SED}}
\end{figure}

Since the amount of radiation reprocessed at every radius is set by the covering fraction ($z/r$) - which is the same for all stars - even the SED in the mid-infrared becomes self-similar (figure \ref{fig:comp_SED}). The scaled model SEDs line up to within 20\% in the mid-infrared at 20-30 micron where we fitted the SEDs. The similarity extends into the far-infrared (within 40\%) and even the near-infrared up to 3 micron (within a factor of 2). The latter is surprising and probably coincidental, given the different nature of the emission mechanism \textit{at that wavelength}: viscous heating for the T Tauri star, optically thin emission for the Herbig star and mostly photospheric flux for the brown dwarf.

\begin{table}
  \title{Disk properties}
  \centering
  \begin{tabular}{lllll}
    \hline \hline
    Parameter  & Brown  & T Tauri & Herbig & Self- \\
               & dwarf  &         & Ae/Be  & similar \\
    \hline
    T$_{\rm 1AU}$ [K]            & 36  & 86   & 185 & 90  \\
    q                & 0.45  & 0.44$^\dagger$  & 0.43 & 0.44  \\
    $\Sigma_{1\rm AU}$ [gr / cm$^2$]  & 0.04 & 0.24 & 0.95  & 0.25  \\
    I$_{\rm 100AU}$ [mJy/AU$^2$]  & 1.1$\cdot 10^{-9}$  & 1.3$\cdot 10^{-6}$  & 4.2$\cdot 10^{-4}$  & 2.9$\cdot 10^{-5}$ \\
    o                           & 3.1  & 2.9  & 2.6   &  2.9 \\
    \hline \hline    
  \end{tabular}
  \caption{Disk properties for our best-fit models. The midplane temperature is fitted by T$_{\rm mp}$(r)=T$_{\rm 1AU}$ r$^{-\rm q}$. The intensity in scattered light by I(r)= I$_{\rm 100AU}$ r$^{-\rm o}$ outwards of 70 AU.
    $^\dagger$ The midplane temperature in T Tauri stars shows a clear break at 2 AU caused by viscous heating (fig \ref{fig:comp_tmid}a), and at smaller radii it is best described as T(r)=125 r$^{-0.93}$.
    \label{tab:disk_properties}}
\end{table}

\subsection{Grain size distributions}\label{sed:discuss_gsd}
Throughout this article, we have assumed that there are no radial gradients in the grain size distribution. Models of grain growth predict that grain sizes decrease with radius \citep[e.g. ][]{2010A&A...516L..14B}, which has also been observed by \cite{2011A&A...529A.105G} and \cite{2011A&A...525A..12B}. Additionally, disks around later type stars are colder and less massive, leading to smaller grain sizes. Modeling the grain size distribution is beyond the scope of this paper, though a recipe for doing so is available \citep{2011A&A...525A..11B}. For now, we will only discuss if we expect to see differences in the grain size distribution in the mid-infrared emitting region between the three different samples.

Because the vertical structure in the mid-infrared region is self-similar, we investigated if this is also true for the grain-size distribution and its planet-forming potential. Grains in protoplanetary disks are in a coagulation-fragmentation equilibrium \citep{1984Icar...60..553W}, driven by turbulence. According to \cite{2011A&A...525A..11B}, the most important parameters that define the size distribution are the turbulent mixing strength, the gas surface density, the temperature, and three parameters that refer to the microphysics of dust: the fragmentation velocity $\rm u_f$, the solid density of dust $\rm \rho_s$ and the size distribution of fragments $\xi$. We assumed the last three to be the same across the three samples.

We find no variations in the turbulent mixing strength (section \ref{sec:obs}). Regions with similar temperature also have a similar surface density (section \ref{sec:self_similar}). The grain size distribution should then also be self-similar. We can see why this is the case by looking at the expression for the maximum grain size for turbulence-driven growth\footnote{which is the relevant regime for particle sizes smaller than a millimeter} - as given by equation (12) of \cite{2009A&A...503L...5B}:
\begin{equation}
 \hat{a}_{\rm max}(r_{\rm T})
 =\frac{\hat{\Sigma}(r_{\rm T})u_{\rm f}^2}{\pi\alpha_{\rm turb} \rho_{\rm s} \hat{c}_{\rm s}^2(r_{\rm T})} 
 =\frac{\Sigma(r) u_{\rm f}^2}{\pi \alpha_{\rm turb} \rho_{\rm s} c_{\rm s}^2(r)} = a_{\rm max}(r),
\end{equation}
which is also self-similar when looking at regions with similar temperature ($\rm r_{\rm T}=r/\sqrt{L_*/L_{\odot}}$). This means that also the grain size distribution in the mid-infrared emitting region is not expected to vary across the stellar mass range.

\subsection{Planet-forming potential}\label{sec:planets}
If the same disk properties are found at a different location in the disk around heavier stars, does this mean that the resulting planetary systems are simply scaled versions of each other?

To compare the later stages of planet formation, we looked at the Hill sphere and feeding zone of a forming planet, to see how much mass is available for its formation, in a way similar to \cite{2007ApJ...669..606R}. The Hill sphere is given by
\begin{equation}
R_{\rm H}(r)=r \sqrt[3]{\frac{m_{\rm P}}{3 M_*}},
\end{equation}
where $m_{\rm P}$ is the mass of the planet. Expressed in the same units that result in self-similar solutions for the vertical structure ($\rm r_{\rm T}=r/\sqrt{\rm L_*/L_{\odot}}$) it becomes
\begin{equation}
\hat{R}_{\rm H}(r_{\rm T})=r/\sqrt{L_*/L_{\odot}} \sqrt[3]{\frac{m_{\rm P}}{3 M_*/\sqrt{L_*/L_{\odot}}}} = 
  \frac{R_{\rm H}(r)}{\sqrt[3]{L_*/L_{\odot}}},
\end{equation}
So the size of the Hill sphere is not self-similar, though its dependence on luminosity is weak. It decreases in size for less luminous stars, despite the lower gravity of the central star. The total mass available in the feeding zone deviates more because it scales with an extra factor $2\pi r_{\rm T}$:
\begin{equation} \begin{split}
\hat{M}_{\rm feed}(r_{\rm T}) &= 2\pi r_{\rm T} \hat{\Sigma}(r_{\rm T})* 2\hat{R}_{\rm H}(r_{\rm T}) \\
    &= 2\pi\Sigma(r)r/\sqrt{L_*/L_\odot} * 2 \frac{R_{\rm H}(r)}{\sqrt[3]{L_*/L_{\odot}}} \\
    &= L_*/L_\odot^{-5/6} M_{\rm feed}(r).
\end{split} \end{equation}
The size of the feeding zone therefore scales almost linearly with luminosity. This means that although the initial conditions for planet formation are the same across the steller mass range, this is not true for the later stages of planet formation, in agreement with \cite{2007ApJ...669..606R}.

\subsection{Dependence on disk parameters}\label{sec:alpha_1d-2}
The method we present here is useful for constraining variations in the turbulent mixing strength, but it is more difficult to constrain its absolute value. The mid-infrared flux with which we fitted our models traces the height of the disk surface, which can also be influenced by other factors than the turbulent mixing strength. However, because our solutions for the mid-infrared emitting region are self-similar, these uncertainties will affect our estimate of the turbulent mixing strength in the same way for all stars, and do not influence our conclusion that it does not vary across the stellar mass range.

In this section we will explore how other parameters can affect our estimate of the turbulent mixing strenght. In particular, we will explore how to keep the turbulent mixing strength fixed at $\alpha_{\rm turb}=0.01$ and fit the three median SEDs by varying these other parameters. As already mentioned in section \ref{sec:tts}, small changes in the grain size distribution also affect the height of the disk surface. Using a flatter slope for this distribution removes small grains from the disk surface, and we can fit the median SEDs with a power-law index of 3.25. Another parameter that directly influences the SED in the mid infrared is the gas-to-dust ratio. A lower ratio leads to a weaker coupling of the dust and hence to flatter disks. We fitted a gas-to-dust ratio of 1 for $\alpha_{\rm turb}=0.01$.

One parameter that we cannot change is the dust mass, because this would be inconsistent with the millimeter photometry. As a consequence, all parameters that influence the millimeter opacity require a refitting of the dust mass. For example, we can reduce the opacity in the upper layers by reducing the carbon content of our dust grains, but this also reduces the opacity at millimeter wavelengths. If we then increase the dust mass to fit the flux at 850 \um, we again increase the opacity in the upper layers. The two effects cancel out, and there is no net decrease in the mid-infrared flux. In a similar fashion, the maximum grain size $a_{\rm max}$ does not influence the estimate of \alfaturb because it decreases the opacity in the upper layers and at millimeter wavelengths simultaneously.

The same is true for parameters that influence the distribution of mass in the outer disk: the outer radius and surface density powerlaw. Although a smaller disk has a smaller covering fraction for the same surface density, a fit to the millimeter data requires a higher surface density, compensating the decrease in covering fraction. The same is true for steeper surface density powerlaws.

\subsection{Spread in disk parameters}\label{sec:spread}

\cite{2005ApJ...628L..65F} have shown that the spread in observed disk colors in the mid infrared is well reproduced by a spread in inclination, accretion and surface layer depletion. However, millimeter observations show that there is also a considerable spread in disk mass \citep{2005ApJ...631.1134A}, which will affect the amount of mid-infrared emission through changing dust surface densities and stronger or weaker dust-gas coupling. We investigated if this is sufficient to explain the observed spread in disk colors, without the need for a spread in surface layer depletion.

\begin{figure}
  \includegraphics[width=\linewidth]{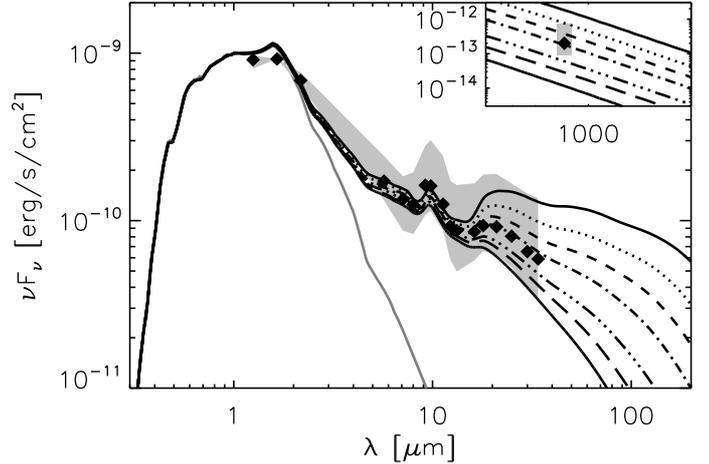}
  \caption[]{Observed median SED of T Tauri stars in Taurus (diamonds) with quartiles (gray area). Overplotted is a series of disk models with varying disk mass, centered around the best fit to the median SED. The dust mass ranges from M$_{\rm dust}=5\e{-4}~M_\odot$ at the top to M$_{\rm dust}=5\e{-6}~M_\odot$ at the bottom, with fixed gas-to-dust ratio. The gray line denotes the stellar photosphere. The inset shows the millimeter regime.
    \label{fig:spread_median}}
\end{figure}

We ran a series of disk models based on the best fit to the Taurus median SED with $\alpha_{\rm turb}=10^{-4}$, varying only the disk mass over two orders of magnitude. The result can be found in figure \ref{fig:spread_median}. Although the spread of this grid reproduces the region between the lower and upper quartile at 20-40 micron well, it is much too large for the quartiles in the millimeter. This means that the real spread in disk mass must be smaller, and therefore cannot be solely responsible for the spread in observed colors. Albeit slightly smaller than previously estimated, a spread in surface layer depletion must still be present\footnote{Which would in our model translate into a spread in turbulent mixing strength or grain size distribution.}.

This result also serves to validate one of the key assumptions of this paper: that a model with median fit parameters is a good representation of the median SED. Because the model SEDs do not overlap (which is also the case when varying most other disk parameters) the median SED is exactly equal to the SED with the median fit parameters.

\section{Conclusions}\label{sec:conclusion}
We have investigated the turbulent mixing strength in protoplanetary disks across the stellar mass range by comparing radiative transfer models with dust settling to median SEDs of three samples of T Tauri stars, Herbig stars and brown dwarfs. Our conclusions are:

\begin{itemize}
\item We find no significant variations of the turbulent mixing strength across the stellar mass range from Herbig stars down to brown dwarfs.
\item We find a relatively low turbulent mixing strength of $\alpha_{\rm turb} = 10^{-4}$ for an MRN-like grain size distribution up to a millimeter and a gas-to-dust ratio of 100, whereas a size distribution with fewer small grains or a gas-to-dust ratio of 1 is required for a higher mixing strength of $\alpha_{\rm turb} = 10^{-2}$.
\item The mid-infrared emitting regions of protoplanetary disks are remarkably self-similar, with a similar surface density, vertical structure of gas and dust, and even grain size distribution. This provides comparable environments for the first stages of planet formation, though the later phases dominated by the planet's Hill sphere are likely different.
\item When viewed at the same distance from the central star, disks around later type stars appear more settled. They have a flatter disk surface as traced in scattered light. However, their gas scale height is much higher.
\item We have constructed a median SED for Herbig stars, analogous to its Taurus and Chameleon I counterparts. This median SED points to high inner-rim temperatures of $\sim$1700 K, and is only poorly fitted with a puffed-up inner rim, but is more consistent with emission from optically thin material from above the rim or from within it.
\end{itemize}

\begin{acknowledgements}
We thank Kees Dullemond for helping with the implementation of self-consistent settling and for discussions during the development of the paper, and Til Birnstiel for useful comments on the manuscript.
The authors would also like to thank Bram Acke, Koen Maaskant and Michiel Min for help and discussion in constructing the Herbig median SED.
We would like to thank the not-so-anonymous referee, Philip Armitage, for his concise and constructive referee report.
  This research project is financially supported by a joint grant from the
  Netherlands Research School for Astronomy (NOVA) and the Netherlands
  Institute for Space Research (SRON). 
\end{acknowledgements}

\bibliographystyle{aa} 
\bibliography{references}

\appendix

\section{Taurus median}\label{app:taurus_median}
As mentioned in section \ref{sec:tts}, we have added the 850 micron flux point to the existing Taurus median from \cite{2006ApJS..165..568F}. Although the original Taurus median from \cite{1999ApJ...527..893D} is based on the same sources and includes millimeter photometry, its coverage is incomplete. It is strongly biased toward brighter - and thus heavier - disks and was 'meant to be indicative rather than definitive'. \cite{2005ApJ...631.1134A} have created a nearly complete census of the Taurus star-forming region at millimeter wavelengths, allowing us to complete the median fluxes at 850 micron. The median flux we derive is $\log(\nu {\rm F}_\nu[\rm erg/s/cm^2])=-12.70$, with the upper and lower quartiles at $-13.05$ and $-12.13$, respectively. These fluxes are roughly an order of magnitude lower than those included in the original median, but are consistent with the average disk mass in Taurus found by \cite{2005ApJ...631.1134A}.

\section{Herbig median}\label{app:herbig_median}
The Herbig median SED is based on the samples of \cite{2009A&A...502L..17A} and \cite{2010ApJ...721..431J}. Because knowing the dust mass is crucial in measuring the degree of settling in the outer disk, we selected only those sources with a millimeter detection. Following \cite{2009A&A...502L..17A}, we also excluded transitional disks, since their mid-infrared SEDs should be explained in the context of gaps and inner holes, rather than dust settling.

The sample consists of 32 sources, and has a complete wavelength coverage in bands, B, V, J, H, K, L, Spitzer IRS longward of 10 micron, IRAS 12, 25 and 60 and millimeter wavelengths. U- and M-band photometry were also included because they lacked only one and two measurements, respectively. The Spitzer spectra were reduced to narrow-band photometry following the same method as in \cite{2006ApJS..165..568F}. At the shortest wavelengths (5.7, 7.1, 8.0, 9.2 \um) the coverage is sometimes incomplete (lacking 5,5,3 and 3 measurements resectively), and was supplemented with ISO data where available. We only included the IRAS 60 micron fluxes, because the 12 and 25 micron regions is covered by Spitzer, and results in the same median fluxes.
Although all sources have millimeter detections, not all are measured at the same wavelength. We therefore constructed one photometric point at 850 \um, and interpolated fluxes according to $\rm \nu F_\nu(mm) = \nu F_\nu(obs)*(\lambda_{obs,mm})^{-4}$, averaging if more than one measurement was available.

The median star in our sample is a Herbig A6 star, with a median effective temperature of 8500 K, luminosity of 21 L$_{\rm \odot}$ and mass of 2 M$_{\rm \odot}$. Because these stars are not in a single star-forming region, we scaled their SEDs to a distance of 140 parsec and the luminosity to the median luminosity before constructing the median. The median including quartiles are given in table \ref{tab:herbig_median}, and are displayed in Figs \ref{fig:herbig_median} and \ref{fig:median_shadow}. This approach results in an SED that has a small spread in photometry at stellar wavelengths, and a similar spread as the Taurus median SED at disk wavelengths. Typical features like the 2 micron bump and silicate features are clearly visible.

\begin{table}
  \centering
  \begin{tabular}{lllll}
    \hline \hline
    band & wavelength  & median $\nu$F$_{\nu}$ & lower quartile & upper quartile \\
         & [$\mu$m] & [erg/s/cm$^2$] & [erg/s/cm$^2$] & [erg/s/cm$^2$] \\
    \hline
    U & 0.36    &  -7.73   &  -7.79   &  -7.70  \\
    B & 0.44    &  -7.46   &  -7.49   &  -7.43  \\
    V & 0.55    &  -7.58   &  -7.61   &  -7.54  \\
    J & 1.23    &  -8.03   &  -8.13   &  -7.87  \\
    H & 1.65    &  -8.11   &  -8.27   &  -7.94  \\
    K & 2.22    &  -8.15   &  -8.36   &  -8.04  \\
    L & 3.77    &  -8.32   &  -8.58   &  -8.16  \\
    M & 4.78    &  -8.47   &  -8.89   &  -8.17  \\
    IRS & 5.70    &  -8.62   &  -9.10   &  -8.43  \\
    IRS & 7.10    &  -8.73   &  -9.19   &  -8.55  \\
    IRS & 8.00    &  -8.63   &  -8.92   &  -8.47  \\
    IRS & 9.20    &  -8.52   &  -8.76   &  -8.24  \\
    IRS & 9.80    &  -8.36   &  -8.67   &  -8.17  \\
    IRS & 11.30    &  -8.50   &  -8.78   &  -8.30  \\
    IRS & 12.30    &  -8.64   &  -8.93   &  -8.41  \\
    IRS & 13.25    &  -8.73   &  -8.93   &  -8.50  \\
    IRS & 16.25    &  -8.64   &  -8.92   &  -8.37  \\
    IRS & 18.00    &  -8.59   &  -8.87   &  -8.29  \\
    IRS & 21.00    &  -8.58   &  -8.91   &  -8.28  \\
    IRS & 25.00    &  -8.64   &  -9.03   &  -8.35  \\
    IRS & 30.00    &  -8.73   &  -9.09   &  -8.44  \\
    IRS & 34.00    &  -8.81   &  -9.18   &  -8.51  \\
    IRAS & 58.61   &  -9.00   &  -9.22   &  -8.51  \\
    mm  & 850.0    &  -11.79  & -12.23   & -11.26  \\
    \hline \hline
  \end{tabular}
  \caption{Median SED for Herbig stars, normalized to a distance of 140 parsec and a luminosity of 21 L$_{\rm \odot}$.
\label{tab:herbig_median}}
\end{table}

\section{The inner regions of Herbig stars}\label{app:inner_rim}
As mentioned in section \ref{sec:herbigs}, the inner regions of Herbig stars deserve special attention in disk modeling. The strong 2 micron bump is a prominent feature in most Herbig Ae and Be stars \citep{2001A&A...365..476M}, and has been explained as the fully illuminated surface of a disk truncated at the dust evaporation radius by \cite{2001ApJ...560..957D}. Because of the large fraction of reprocessed light in the near-infrared, this would require the rim to be 'puffed up', a natural cause of its high temperature down to the midplane \citep{2001A&A...371..186N}. 

However, \cite{2006ApJ...636..348V} showed that a rim in hydrostatic equilibrium does not puff up far enough to explain the near-infrared fluxes. To fit the SED, the inner rim has to be puffed up beyond hydrostatic equilibrium by a factor 2-3 \citep{2010A&A...516A..48V,2009A&A...502L..17A}. Other explanations have also been proposed, ranging from dust halos \citep{2006ApJ...636..348V,2011A&A...528A..91V} to optically thin gas or dust within the inner rim \citep{2008ApJ...676..490K,2008ApJ...689..513T}, as well as theoretical motivations for an increased scale height in the inner rim \citep{2011MNRAS.412..711T} or the presence of such a halo \citep{2011A&A...531A..80K}. For the context of this paper, it is important whether or not the added emission in the inner disk influences the geometry and emission of the outer disk, and could affect our estimate of the turbulent mixing strength. A puffed-up inner rim casts a shadow on the outer disk, decreasing the flux at longer wavelengths \citep{2009A&A...502L..17A}, whereas material within the inner rim or a halo does not cast a shadow \citep{2010A&A...512A..11M}. 

\begin{figure}
  \includegraphics[width=\linewidth]{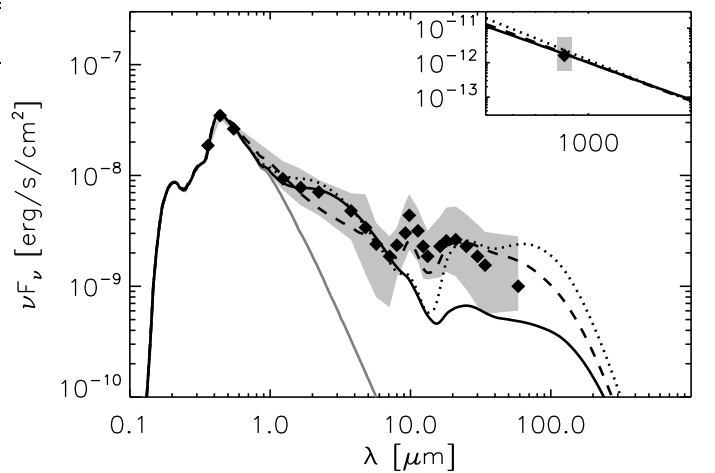}
  \caption[]{Observed median SED of Herbig stars (diamonds) with quartiles (gray area). Overplotted is a disk model with a puffed-up inner rim with a scale height 2.5 times the pressure scale height and turbulent mixing strength of $\alpha=10^{-2}$ (solid line). Also plotted are a model with the same inner rim and a modified grain size distribution with a power-law index of 4.0 (dotted line) and a model without a puffed up inner rim, but with millimeter sized grains within the dust evaporation radius (dashed line). The gray line denotes the stellar photosphere. The inset shows the millimeter regime.
    \label{fig:median_shadow}}
\end{figure}

Figure \ref{fig:median_shadow} shows these alternative inner disk geometries. When we increased the scale height in the inner rim by a factor 2.5 to fit the near infrared part of the SED (Fig \ref{fig:median_shadow}, solid line), it cast a shadow over almost the entire outer disk, and we were unable to fit the far-infrared flux even with a turbulent mixing strength of $\alpha=10^{-2}$ or higher. We were able fit the far infrared flux by increasing the amount of small grains, which we did by increasing the slope of the grain size distribution to q=4.0 (dotted line). However, the strong effect of the inner rim shadow remains, supressing the mid-infrared flux. We tried several prescriptions to modify the inner rim structure, such as rounding it off, but they did not provide enough mid infrared flux. The reason why the puffing up the inner rim does not work well in our model - in contrast to e.g. \cite{2009A&A...502L..17A} and \cite{2011MNRAS.412..711T} - is that the median SED requires additional flux at 2 micron - whereas previous work focused mainly on the 8 micron region. This requires a hot inner rim, 1500-1700 K, which is consistent with observations and dust evaporation, but which does not produce enough MIR emission.

Alternatively, dust or gas within the inner rim can also contribute to the near-infrared flux. Additional components originating in that region have been observed for a number of Herbig stars, such as MWC 147 \citep{2008ApJ...676..490K,2010ApJ...724L...5B}, MWC 275 and AB Aurigae \citep{2008ApJ...689..513T,2008ApJ...677L..51T},  HD 163296 \citep{2010A&A...511A..74B} and HR599 \citep{2011A&A...531A..84B}. Material within the inner rim can increase the near-infrared flux if it generates extra luminosity from accretion \citep{2008ApJ...676..490K} or if it captures additional radiation that would otherwise not reach the disk.

For the second option, the material needs to extend sufficiently close to the star, such that starlight is captured that crosses the midplane and would normally not be reprocessed by the disk. We modeled this by extending the largest grain component in our model down to two stellar radii, or 0.02 AU (Fig \ref{fig:median_shadow}, dashed line). Because even the largest grains in our model at this location become extremely hot, we were unable to fit the shape of the near-infrared SED this way, but it does capture roughly sufficient energy at shorter wavelengths. Because this solution to the near-infrared flux problem is similar to a halo in the sense that it captures additional energy, we did not explore it in more detail. We limited ourselves to modeling a halo, where we note that this could also represent a solution with material within the inner disk.

\end{document}